\documentclass{ieeeaccess}
\usepackage{cite}
\usepackage{amsmath,amssymb,amsfonts}
\usepackage{amsthm}
\usepackage{mathtools}
\usepackage{braket}
\usepackage{lipsum}

\usepackage{caption}
\usepackage{subcaption}
\usepackage{multirow}
\theoremstyle{definition}

\DeclareMathOperator*{\argmin}{arg\,min}

\usepackage{color}
\usepackage{algorithm}
\usepackage{algpseudocode}
\algnewcommand{\StateIndent}[2]{\Statex \hskip#1em #2}

\usepackage[pdftex]{graphicx}
\usepackage{textcomp}

\usepackage{tikz}
\usetikzlibrary{quantikz2}
\usetikzlibrary{backgrounds,fit,decorations.pathreplacing,calc}
\NewSpotColorSpace{PANTONE}
\AddSpotColor{PANTONE} {PANTONE3015C} {PANTONE\SpotSpace 3015\SpotSpace C} {1 0.3 0 0.2}
\SetPageColorSpace{PANTONE}%

\def\BibTeX{{\rm B\kern-.05em{\sc i\kern-.025em b}\kern-.08em
    T\kern-.1667em\lower.7ex\hbox{E}\kern-.125emX}}
\begin{document}
\history{Date of publication xxxx 00, 0000, date of current version xxxx 00, 0000.}
\doi{10.1109/TQE.2020.DOI}

\title{Compressed space quantum approximate optimization algorithm for constrained combinatorial optimization}
\author{\uppercase{Tatsuhiko Shirai}\authorrefmark{1}, \IEEEmembership{Member, IEEE}
\uppercase{and Nozomu Togawa\authorrefmark{2}},
\IEEEmembership{Member, IEEE}
}
\address[1]{Waseda Institute for Advanced Study, Waseda University}
\address[2]{Department of Computer Science and Communications Engineering, Waseda University}
%\tfootnote{This work was supported in part by JST CREST Grant Number JPMJCR19K4, Japan and KAKENHI Grant No.21K03391.
%}

\markboth
{Author \headeretal: Preparation of Papers for IEEE Transactions on Quantum Engineering}
{Author \headeretal: Preparation of Papers for IEEE Transactions on Quantum Engineering}

\corresp{Corresponding author: Tatsuhiko Shirai (email: tatsuhiko.shirai@aoni.waseda.jp).}

\begin{abstract}
Combinatorial optimization is a promising area for achieving quantum speedup.
The quantum approximate optimization algorithm (QAOA) is designed to search for low-energy states of the Ising model, which correspond to near-optimal solutions of combinatorial optimization problems (COPs).
However, effectively dealing with constraints of COPs remains a significant challenge. 
Existing methods, such as tailoring mixing operators, are typically limited to specific constraint types, like one-hot constraints.
To address these limitations, we introduce a method for engineering a compressed space that represents the feasible solution space with fewer qubits than the original.
Our approach includes a scalable technique for determining the unitary transformation between the compressed and original spaces on gate-based quantum computers.
We then propose compressed space QAOA, which seeks near-optimal solutions within this reduced space, while utilizing the Ising model formulated in the original Hilbert space. 
Experimental results on a quantum simulator demonstrate the effectiveness of our method in solving various constrained COPs.
%Combinatorial optimization is a promising area for achieving quantum speedup, but incorporating constraints remains challenging.
%Constraints can reduce algorithm performance by creating large energy barriers between the low-energy feasible solution space and the high-energy infeasible solution space. 
%Existing methods, such as tailoring mixer terms, are limited to specific constraint types, like the one-hot constraint.
%To address these limitations, we introduce a method for engineering a compressed space that represents the feasible solution space with fewer qubits than the original.
%Our approach includes a scalable technique for determining the unitary transformation between the compressed and original spaces on gate-based quantum computers. 
%We then propose compressed space quantum approximate optimization algorithm (CS-QAOA), which seeks near-optimal solutions within this reduced space.
%Experimental results on a quantum simulator demonstrate the effectiveness of our method for solving various constrained combinatorial optimization problems.
%The figure below shows the simulation result of applying compressed QAOA to quadratic knapsack problems (QKP) with an inequality constraint.
%Compressed QAOA outperforms conventional QAOA.
%This work provides a new framework to solve various types of constrained combinatorial optimization problems.
\end{abstract}


\titlepgskip=-15pt

\maketitle

\section{Introduction}
\label{sec:introduction}
\PARstart{C}{ombinatorial} optimization has gained much attention as a potential area for quantum speedup with a wide range of applications including logistics, communications, and finance.
It involves finding the optimal solution that minimizes or maximizes the objective function while satisfying constraints.
Various methods have been developed to map these problems into the Ising models, enabling the use of quantum devices~\cite{lucas2014Ising,glover2018tutorial,shirai2023spin,montanez2024unbalanced}.
The quantum approximate optimization algorithm (QAOA)~\cite{farhi2014quantum, kostas2024review}, a digitized version of quantum annealing~\cite{kadowaki1998quantum}, is designed to search for low-energy states of the Ising model, which correspond to near-optimal solutions of combinatorial optimization problems (COPs). 

Although the QAOA has been extensively studied for unconstrained problems, such as the Max-cut problem~\cite{farhi2014quantum, wang2018quantum, guerreschi2019qaoa, zhou2020quantum, bravyi2020obstacles, wurtz2021maxcut}, Grover's unstructured search~\cite{jiang2017near}, and others~\cite{wauters2020polynomial, farhi2022quantum, rusian2024evidence}, solving constrained COPs remains both practically important and challenging.
%Constraints of a COP separate the configuration space into a feasible solution space with low energies and an infeasible solution space with high energies.
Although the penalty method~\cite{lucas2014Ising} provides a general framework for incorporating constraints, it complicates the energy landscape, resulting in performance degradation.
To improve performance, several approaches have been developed.
Designing mixer terms has been shown to explore solutions within the feasible solution space~\cite{hadfield2019quantum,wang2020xy,bartschi2020grover}, where feasible solutions are those that satisfy the constraints.
Space-efficient embedding techniques reduce the number of qubits required to formulate a COP, thereby decreasing the overall search space~\cite{tabi2020quantum,fuchs2021efficient,glos2022space}.
Post-processing techniques to find feasible solutions enhance QAOA performance~\cite{shirai2024post}.
However, these approaches are generally limited to specific types of constraints, such as one-hot constraints.

%Transitions between the feasible solutions are severely suppressed by the energy barriers of infeasible solutions.
%Reaching the (near)-optimal solution from a local minimum solution is difficult due to the complex energy landscape.

\begin{figure*}[h]
    \centering
    \includegraphics[width=\linewidth]{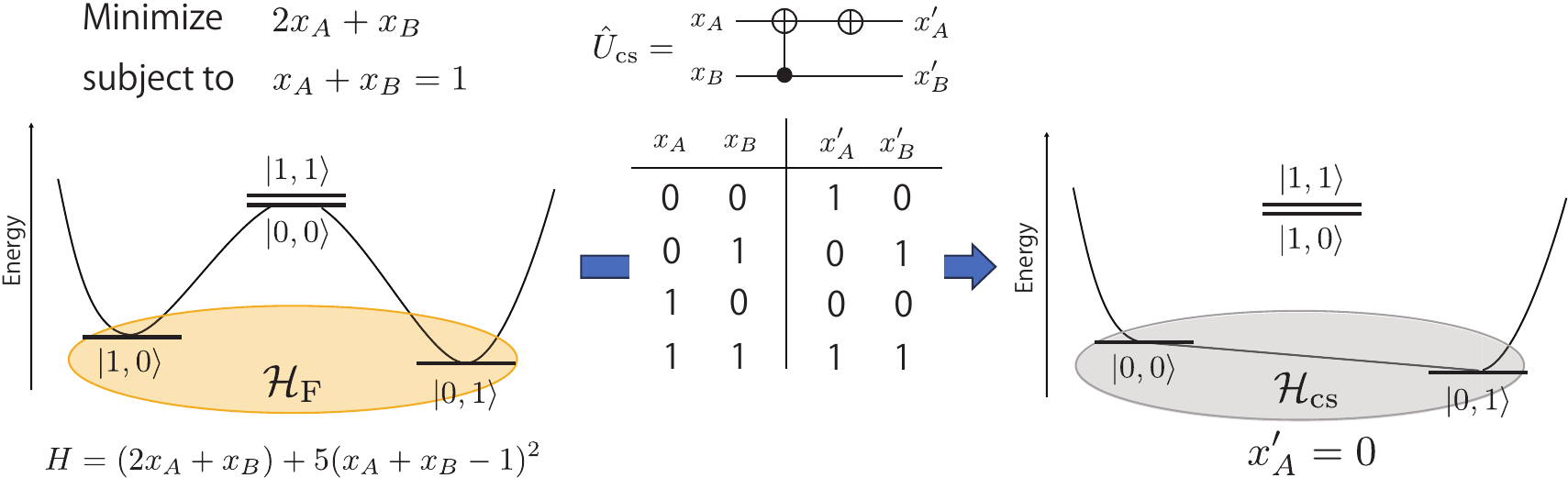}
    \caption{Schematic picture of a compressed space $\mathcal{H}_\mathrm{cs}$ for a simple constrained COP with two binary variables $x_A, x_B \in \{0, 1\}$.
    The feasible solution space $\mathcal{H}_\mathrm{F}$ is mapped to $\mathcal{H}_\mathrm{cs}$ by the unitary transformation $\hat{U}_\mathrm{cs}$.
    Local minimum disappears in the compressed space.
    }
    \label{fig:schematic}
\end{figure*}

An earlier approach attempted to address the issue of restricted applicability by formulating constrained COPs as Lagrange-duality problems~\cite{le2024solving}.
In this work, we present an alternative solution that smoothly transforms the energy landscape, drawing inspiration from simulated annealing~\cite{shirai2022multi, jimbo2024gpu} and quantum annealing~\cite{fujii2023eigenvalue}.
Specifically, we develop a method to engineer a compressed space, which expresses the feasible solution space with fewer qubits than the original configuration space~\cite{botelho2022error}.
The following example illustrates how compressed space engineering enhances the QAOA performance.
Fig.~\ref{fig:schematic} shows a schematic of the compressed space formulation for a simple constrained COP with two binary variables $x_A, x_B \in \{0, 1\}$.
The objective is to minimize the function $2x_A+x_B$, subject to the constraint $x_A+x_B=1$.
The optimal solution is $(x_A, x_B)=(0, 1)$.
The left panel of Fig.~\ref{fig:schematic} shows the energy landscape constructed using the penalty method with two variables $x_A$ and $x_B$\footnote{The energy function is called quadratic unconstrained binary optimization, which is equivalent to the Ising model.}, where the first term describes an objective function and the second term represents the constraint.
A constraint coefficient $5$ ensures that the feasible solution states $\ket{0,1}$ and $\ket{1,0}$ have low energies.
The ground state $\ket{0,1}$ corresponds to the optimal solution, whereas $\ket{1,0}$ is a local minimum, making it difficult to reach the optimal solution via single-variable flips.
Now we introduce a compressed space with a single binary variable $x'_B \in \{0, 1\}$, where $x'_A$ is fixed to zero.
This reduces the dimension of the space to match that of the feasible solution space.
As shown in Fig.~\ref{fig:schematic}, a unitary operator maps the feasible solution space to the compressed space, where the states $\ket{0,1}$ and $\ket{1,0}$ correspond to $x'_B=1$ and $0$, respectively.
Importantly, the local minimum solution disappears in the compressed space.
%The idea of transforming energy landscape smoothly has been demonstrated to improve the performance of simulated annealing~\cite{shirai2022multi} and to mitigate the small energy gap during quantum annealing~\cite{fujii2023eigenvalue}.

Building on the observation, we propose the compressed space QAOA (CS-QAOA).
The CS-QAOA can explore the (near)-optimal solutions within the compressed space, while utilizing the Ising model formulated in the original Hilbert space.
We show two types of methods for engineering the compressed space.
One is the deterministic method for specific types of constraints called the one-hot constraint and the parity constraint~\cite{wolfgang2015quantum}, whereas the other is the variational method aiming to extend the applicability range of the CS-QAOA.
Throughout the application of the CS-QAOA to Max-3 cut problems and quadratic assignment problems (QAPs) with one-hot constraints and quadratic knapsack problems (QKPs) with an inequality constraint, we demonstrate its effectiveness compared to the conventional QAOA.
Furthermore, we examine the impact of noise on its performance and clarify the limitations of its applicability on current quantum devices.
%We also clarify the limitation of its application range.

The remaining part of this paper is organized as follows:
Section~\ref{sec:preliminary} provides the settings and preliminaries. Section~\ref{sec:cs-qaoa} proposes the CS-QAOA.
Section~\ref{sec:case} applies the proposed algorithm to several COPs and demonstrates its effectiveness.
Finally, Section~\ref{sec:conclusion} concludes this paper.

\section{Preliminaries}
\label{sec:preliminary}
This paper considers the following COPs:
For $x_i \in \{0, 1\}$ where $i \in \{1,2, \ldots, N\} = V$, find $\argmin f(\{x_i\}_{i \in V})$ under constraints such that for $k \in M$, $g_k (\{x_i\}_{i \in V_k}) \in [\ell_k, u_k]$.
Here, $f(\{x_i\}_{i \in V})\in \mathbb{R}$ is an objective function, $g_k (\{x_i\}_{i \in V_k})$ and $\ell_k \leq u_k$ are integers, $V_k \subseteq V$, and $M$ is a set of constraints. 
The set of feasible solutions is denoted by
\begin{equation}
    F = \{ \{x_i\}_{i\in V} | \forall k \in M, g_k (\{x_i\}_{i \in V_k}) \in [\ell_k, u_k] \}.
    \label{eq:feasible}
\end{equation}
The number of feasible solutions is given by $|F|$, where $|A|$ denotes the cardinality of the set $A$.

The Ising model is a common input for quantum devices.
The Hamiltonian acts on the Hilbert space $\mathcal{H}$ with $N$ spins (qubits) and is given by
\begin{equation}
\hat{H}_\mathrm{Ising} = \sum_{i=1}^N \sum_{j=i+1}^N J_{ij} \hat{\sigma}_i^z \hat{\sigma}_j^z + \sum_{i=1}^N h_i \hat{\sigma}_i^z +H_0,
\label{eq:Ising}
\end{equation}
where $\{\hat{\sigma}_i^\alpha\}_{\alpha \in \{x,y,z\}}$ is the Pauli spin operator acting on site $i \in V$.
$J_{ij} \in \mathbb{R}$ denotes the interaction between spins $i$ and $j$, $h_i \in \mathbb{R}$ denotes the bias on spin~$i$, and $H_0 \in \mathbb{R}$.
Each solution of a COP is mapped to the eigenstate of $\hat{H}_\mathrm{Ising}$ denoted by $\ket{x} = \bigotimes_{i\in V} \ket{x_i}$.
Throughout this study, we use the notation that $\hat{\sigma}_i^z \ket{x_i} = (2x_i - 1) \ket{x_i}$, where $x_i \in \{0, 1\}$.
Then the feasible solution space $\mathcal{H}_\mathrm{F} \subseteq \mathcal{H}$ is spanned by the eigenstates of $\hat{H}_\mathrm{Ising}$ that satisfy the constraints of the COP, i.e.
\begin{equation}
    \mathcal{H}_\mathrm{F} = \mathrm{Span} (\{ \ket{x} | \{x_i\}_{i\in V} \in F\}).
\end{equation}
The dimension of $\mathcal{H}_\mathrm{F}$ is equal to $|F|$ (i.e., $\mathrm{dim}\mathcal{H}_\mathrm{F}=|F|$).

The QAOA utilizes both quantum and classical devices to find low-energy states of the Ising model~\cite{farhi2014quantum}.
The quantum device generates a variational state $\ket{\vec{\beta}, \vec{\gamma}} \in \mathcal{H}$ using a quantum circuit with real-valued parameters $\vec{\beta} = (\beta_1, \ldots, \beta_p)$ and $\vec{\gamma}= (\gamma_1, \ldots, \gamma_p)$, where $p \in \mathbb{N}$ denotes the number of QAOA layers.
On the other hand, the classical computer calculates the energy expectation value $E(\vec{\beta}, \vec{\gamma}) = \bra{\vec{\beta}, \vec{\gamma}} \hat{H}_\mathrm{Ising} \ket{\vec{\beta}, \vec{\gamma}}$ and updates parameters $\vec{\beta}$ and $\vec{\gamma}$ to lower the energy using a classical optimizer such as the Powell method and the biconjugate gradient method.
This process is repeated until the parameters $\vec{\beta}$ and $\vec{\gamma}$ converge, with the converged values denoted by $\vec{\beta}^*$ and $\vec{\gamma}^*$.

Fig.~\ref{fig:qaoa}~(a) depicts the quantum circuit of the conventional QAOA.
The quantum circuit consists of the phase operator $\hat{U}_\mathrm{C} (\gamma_i)$ and the mixing operator $\hat{U}_\mathrm{M}(\beta_i)$ in alternation, where $i \in \{1, \ldots, p\}$.
The variational state $\ket{\vec{\beta}, \vec{\gamma}}$ is given by
\begin{equation}
    \ket{\vec{\beta}, \vec{\gamma}} = \left( \prod_{i=1}^p \hat{U}_\mathrm{M} (\beta_{p+1-i}) \hat{U}_\mathrm{C} (\gamma_{p+1-i}) \right) \ket{\psi_\mathrm{ini}},
\end{equation}
where $\ket{\psi_\mathrm{ini}}$ is an initial state.
The phase operator is given by the Ising model as $\hat{U}_\mathrm{C} (\gamma)= \exp (-i\gamma \hat{H}_\mathrm{Ising})$.
We consider two types of mixing operators: X mixer and XY mixer.
The X mixer and the XY mixer are given by $\hat{U}_\mathrm{X}(\beta)= \exp (-i\beta \sum_{j\in V} \hat{\sigma}_j^x)$ and $\hat{U}_\mathrm{XY}(\beta)= \exp [-i \beta \sum_{j<k}^N(\hat{\sigma}_j^x \hat{\sigma}_k^x + \hat{\sigma}_j^y \hat{\sigma}_k^y)/2]$, respectively~\cite{wang2020xy}.
It is noted that compiling the XY mixer on gate-based quantum computers requires a Trotterization~\cite{trotter1959product,hatano2005finding} or a partitioning into disjoint pairs $\hat{U}_\mathrm{XY}(\beta) \approx \prod_{j<k}^N \exp [-i \beta (\hat{\sigma}_j^x \hat{\sigma}_k^x + \hat{\sigma}_j^y \hat{\sigma}_k^y)/2]$~\cite{hadfield2019quantum}.
%Here, $V_X$ and $E_{XY}$ are the set of nodes and edges, respectively.

\begin{figure}
\begin{minipage}[]{0.45\linewidth}
%\centering
\subcaption{}
\scalebox{0.7}{
\begin{quantikz}
   \lstick[wires = 4]{$\ket{\psi_\mathrm{ini}}$}  &\gate[wires=3]{\hat{U}_\mathrm{C} (\gamma_i)} \gategroup[3,steps=2,style={inner sep=6pt}]{$\times p$} &\gate[wires=3]{\hat{U}_\mathrm{M} (\beta_i)}&\rstick[wires = 4]{$\ket{\vec{\beta}, \vec{\gamma}}$}\\
      &{}& &{}\\
      &{}& &{}
\end{quantikz}\\
%\begin{quantikz}
%   \lstick[wires = 4]{$\ket{+^n}$}  &\gate[wires=3]{\hat{U}_C (\gamma_i)} \gategroup[3,steps=2,style={inner sep=6pt}]{$\times p$} &\gate[wires=1]{\hat{U}_X (\beta_i)}&\rstick[wires = 4]{$\ket{\vec{\beta}, \vec{\gamma}}$}\\
%      &{}& \gate[wires=1]{\hat{U}_X (\beta_i)} &{}\\
%      &{}& \gate[wires=1]{\hat{U}_X (\beta_i)} &{}
%\end{quantikz}
}
\end{minipage}\\
\begin{minipage}[]{0.45\linewidth}
%\centering
\subcaption{}
\scalebox{0.7}{
\begin{quantikz}
    \lstick{$\ket{0^{N-m}}$}  &\gate[wires=3,style=red!20]{\hat{U}_\mathrm{cs}^\dagger} &\gate[wires=3]{\hat{U}_\mathrm{C} (\gamma_i)} \gategroup[3,steps=5,style={inner sep=6pt}]{$\times p$} &\gate[wires=3,style=red!20]{\hat{U}_\mathrm{cs}}&\meter{}&\wireoverride{} \ket{0^{N-m}}&\gate[wires=3,style=red!20]{\hat{U}_\mathrm{cs}^\dagger}&\rstick[wires = 4]{$\ket{\vec{\beta}, \vec{\gamma}}$}\\
    \lstick[wires = 3]{$\ket{+^m}$}  &{}&{}&{}&\gate[wires=2]{\hat{U}_\mathrm{X} (\beta_i)}& {} & {} &{}\\
     &{}&{}&{}&{}& {} &{} &{}
\end{quantikz}
}
\end{minipage}
\caption{Quantum circuits of (a)~the conventional QAOA and (b)~the CS-QAOA to generate a variational state $\ket{\vec{\beta},\vec{\gamma}}$. $p$ and $m$ denote the number of QAOA layers and the number of qubits in the compressed space, respectively.}
\label{fig:qaoa}
\end{figure}

\section{compressed space QAOA}
\label{sec:cs-qaoa}
We introduce a compressed space $\mathcal{H}_\mathrm{cs}$ following~\cite{botelho2022error}.
%Consider a feasible solution space $\mathcal{H}_\mathrm{F} \subseteq \mathcal{H}$, which is spanned by the eigenstates of $H_\mathrm{Ising}$ that satisfies the constraints of the combinatorial optimization problem.
The compressed space $\mathcal{H}_\mathrm{cs}$ is a Hilbert space with $m$ qubits, where $m$ is determined by the dimension of the feasible solution space so that $\mathrm{dim} \mathcal{H}_\mathrm{cs} =2^m \geq |F|$ and $m < N$.
The mapping between $\mathcal{H}_\mathrm{F}$ and $\mathcal{H}_\mathrm{cs}$ is given by a unitary operator $\hat{U}_\mathrm{cs}$:
\begin{equation}
\forall \ket{\psi} \in \mathcal{H}_\mathrm{F}, \exists \ket{\phi} \in \mathcal{H}_\mathrm{cs} \text{ s.t. } \hat{U}_\mathrm{cs} \ket{\psi} = \ket{0^{N-m}} \ket{\phi}.
\label{eq:compress}
\end{equation}
Fig.~\ref{fig:schematic} illustrates the compressed space formulation when $N=2$ and $m=1$.

Fig.~\ref{fig:qaoa}~(b) gives the quantum circuit of the CS-QAOA to generate a variational state $\ket{\vec{\beta}, \vec{\gamma}} \in \mathcal{H}$.
The quantum circuit consists of $\hat{U}_\mathrm{C}(\gamma_i)$ and $\hat{U}_\mathrm{X}(\beta_i)$ for $i \in \{1, \ldots, p\}$.
The X~mixers act to the qubits in the compressed space.
The initial state is set to $\ket{0^{N-m}} \ket{+^m}$, where $\ket{+}=(\ket{0}+\ket{1})/\sqrt{2}$.
The key distinction of the CS-QAOA from the conventional QAOA lies in the inclusion of the unitary operator $\hat{U}_\mathrm{cs}$ and its Hermite conjugate $\hat{U}_\mathrm{cs}^\dagger$ that give a mapping between the feasible solution space and the compressed space.
The quantum circuit includes mid-circuit measurements of $N-m$ qubits, which do not belong to the compressed space.
The $N-m$ qubits are reset to $0$~states only when the measurements result in all zeros.
Otherwise, the calculation is discarded.
%It is noted that discarding only occurs in the presence of errors caused by imperfections in $\hat{U}_\mathrm{cs}$ or external noise.

The CS-QAOA is regarded as a QAOA with a tailored mixer designed to explore the feasible solution space~\cite{wang2020xy}, fitting within the framework of the quantum alternating operator ansatz~\cite{hadfield2019quantum}.
In the example of Fig.~\ref{fig:schematic}, $\hat{U}_\mathrm{cs}^\dagger \ket{0} \ket{+} = (\ket{01}+\ket{10})/\sqrt{2} \eqqcolon \ket{W_V}$, which is a uniform superposition state of the feasible solutions, and $\hat{U}_\mathrm{cs} (\hat{I} \otimes \hat{U}_X(\beta)) \hat{U}_\mathrm{cs}^\dagger$ is equivalent to $\hat{U}_\mathrm{XY}(\beta)$ in the feasible solution space.
Here, $\ket{W_V}$ is the $W$-state~\cite{dur2000three} for qubits in $V$ and $\hat{I}$ denotes the identity operator.
Notably, the CS-QAOA does not explicitly require the preparation of the superposition state of feasible solutions and the design of the tailored mixer as long as a unitary operator $\hat{U}_\mathrm{cs}$ can be identified.

The CS-QAOA is also related to the method with space-efficient embedding~\cite{tabi2020quantum,fuchs2021efficient,glos2022space}.
Namely, $\hat{U}_\mathrm{cs}^\dagger \hat{U}_\mathrm{C}(\gamma) \hat{U}_\mathrm{cs}$ non-trivially acts on the qubits only in a compressed space, i.e., $\hat{U}_\mathrm{cs}^\dagger \hat{U}_\mathrm{C}(\gamma) \hat{U}_\mathrm{cs} = \hat{I}^{N-m} \otimes \exp(-i\gamma \hat{H}_\mathrm{se})$ where $\hat{H}_\mathrm{se} = \bra{0^{N-m}}\hat{U}_\mathrm{cs}^\dagger \hat{H}_\mathrm{Ising} \hat{U}_\mathrm{cs} \ket{0^{N-m}}$.
However, directly obtaining $\hat{H}_\mathrm{se}$ is generally challenging, except in the case of one-hot constraint.

In contrast to previous approaches, the introduction of compression unitary operators $\hat{U}_\mathrm{cs}$ provides a flexible means of engineering mixers, including the variational method discussed in the following.

\subsection{Engineering of compressed space}
\label{sec:experiment}
Here, we provide methods for engineering the unitary operator $\hat{U}_\mathrm{cs}$ for the CS-QAOA.
First, we explain the deterministic and variational methods to consider a feasible solution space with a single constraint.
Then, we propose an algorithm that leverages these methods to handle multiple constraints in \eqref{eq:feasible}.

\subsubsection{Deterministic method}
\begin{figure}
\begin{minipage}[]{0.45\linewidth}
%\centering
\subcaption{One-hot constraint}
\scalebox{0.7}{
\begin{quantikz}
    \lstick[4]{$\ket{\psi}\in \mathcal{H}_\mathrm{F}$}&{}&{}&\swap{2}&\targ{}&\targ{}&\rstick[1]{$\ket{0}$}\\
    {}&\targ{}&{}&{}&\ctrl{-1}&\swap{1}&\rstick[1]{$\ket{0}$}\\
    {}&{}&\ctrl{1}&\targX{}&{}&\targX{}&\rstick[2]{$\ket{\phi} \in \mathcal{H}_\mathrm{cs}$}\\
    {}&\ctrl{-2}&\targ{}&\ctrl{-1}&{}&{}&
\end{quantikz}\quad
}
\end{minipage}\\
\begin{minipage}[]{0.45\linewidth}
%\centering
\subcaption{Parity constraint}
\scalebox{0.7}{
\begin{quantikz}
    \lstick[4]{$\ket{\psi}\in \mathcal{H}_\mathrm{F}$}&\targ{}&\targ{}&\targ{}&\rstick[1]{$\ket{0}$}\\
    {}&\ctrl{-1}&{}&{}&\rstick[3]{$\ket{\phi} \in \mathcal{H}_\mathrm{cs}$}\\
    {}&{}&\ctrl{-2}&{}&{}\\
    {}&{}&{}&\ctrl{-3}&{}\\
%\end{quantikz}\\
%\begin{quantikz}
    \lstick[4]{$\ket{\psi}\in \mathcal{H}_\mathrm{F}$}&\targ{}&\targ{}&\targ{}&\targ{}&\rstick[1]{$\ket{0}$}\\
    {}&{}&\ctrl{-1}&{}&{}&\rstick[3]{$\ket{\phi} \in \mathcal{H}_\mathrm{cs}$}\\
    {}&{}&{}&\ctrl{-2}&{}&{}\\
    {}&{}&{}&{}&\ctrl{-3}&{}
\end{quantikz}
}
\end{minipage}
\caption{Quantum circuits for generating the compressed space for (a)~one-hot constraint $\hat{U}_{\mathrm{binary},V} \ket{\psi}=\ket{0}^{N-m} \ket{\phi}$ and (b)~parity constraint $\hat{U}_{\mathrm{even (odd)},V} \ket{\psi}=\ket{0}^{N-m} \ket{\phi}$, where $\ket{\psi} \in \mathcal{H}_\mathrm{F}$, $\ket{\phi} \in \mathcal{H}_\mathrm{cs}$, and $N=4$.
Fig.~\ref{fig:deterministic}~(b) (upper) and (lower) correspond to the quantum circuit for the even and odd parity constraints, respectively.}
\label{fig:deterministic}
\end{figure}

The compressed space can be created for specific types of constraint.
First, we consider a one-hot constraint: $\sum_{i \in V} x_i = 1$.
%\begin{equation}
%    \sum_{i \in V} x_i = 1.
%\end{equation}
This constraint requires that one qubit among $N$ qubits takes the value one, making $|F|= N$.
For this case, a unitary operator denoted by $\hat{U}_{\mathrm{binary},V}$ that converts a one-hot representation to a binary representation~\cite{sawaya2020resource} creates a compressed space with $m$ qubits, where $m=\lceil \log_2 N \rceil$.
Here, $\lceil x \rceil = \min \{m \in \mathbb{Z}| m \geq x\}$ is the ceiling function.
%We denote the unitary operator by $\hat{U}_{\mathrm{binary},V}$, which maps a state in $\mathcal{H}_\mathrm{F}$ to a state in the compressed space.
%It acts on $\ket{\phi} \in \mathcal{H}_\mathrm{cs}$ as $\hat{U}_{\mathrm{binary},V}^\dagger \ket{0^{N-m}} \ket{\phi} \in \mathcal{H}_\mathrm{F}$.
%\begin{equation}
%    \hat{U}_{\mathrm{binary},V}^\dagger \ket{0^{N-m}} \ket{\phi} \in \mathcal{H}_\mathrm{F}
%\end{equation}
Fig.~\ref{fig:schematic} and Fig.~\ref{fig:deterministic}~(a) provide examples of $\hat{U}_{\mathrm{binary},V}$ for $N=2$ and $4$, respectively.
For other values of $N$, including the cases where $N$ is not a power of two, refer to~\cite{sawaya2020resource}.

Next, we consider a parity constraint that requires the sum of $x_i$ for $i \in V$ to be either even or odd.
We refer to these as the even-parity constraint and the odd-parity constraint, respectively.
For both parity constraints, $|F|= 2^{N-1}$, meaning each constraint generates a compressed space with $N-1$ qubits.
The unitary operators for generating the compressed space are given as 
%\begin{equation}
%    \hat{U}_{\mathrm{even (odd)},V}^\dagger \ket{0} \ket{\phi} \in \mathcal{H}_\mathrm{F},
%\end{equation}
\begin{equation}
    \hat{U}_{\mathrm{even},V}= \prod_{i=2}^{N} \mathrm{CX}_{i,1}, \quad
    \hat{U}_{\mathrm{odd},V}= \hat{\sigma}_1^x \prod_{i=2}^{N} \mathrm{CX}_{i,1}.
\end{equation}
%where $\hat{U}_{\mathrm{even (odd)},V}^\dagger \ket{0} \ket{\phi} \in \mathcal{H}_\mathrm{F}$.
Here, $\mathrm{CX}_{i,j}$ is the CNOT gate with $i$-th qubit being the control qubit and $j$-th qubit being the target qubit.
Fig.~\ref{fig:deterministic}~(b) provide implementations of $\hat{U}_{\mathrm{even},V}$ and $\hat{U}_{\mathrm{odd},V}$ for $N=4$.

It is noted that discarding through the mid-circuit measurements in Fig.~\ref{fig:qaoa}~(b) does not occur when $\hat{U}_{\mathrm{binary},V}$ is implemented for the one-hot constraint, or when $\hat{U}_{\mathrm{even(odd)},V}$ is used for the parity constraint as $\hat{U}_\mathrm{cs}$, provided that there are no gate errors.

\subsubsection{Variational method}
The variational method extends the applicability range of the CS-QAOA.
First, we introduce a compressed-space Hamiltonian $\hat{H}_\mathrm{cs}$.
This Hamiltonian satisfies the following conditions:
(i) $\hat{H}_\mathrm{cs}: \mathcal{H} \to \mathcal{H}$,
(ii) diagonal in the computational basis, (iii) all the eigenvalues of the feasible solutions are smaller than those of any infeasible solutions, and (iv) the eigenvalues are not degenerate.

The compressed-space Hamiltonian can be constructed for various types of constraints.
Let us consider a constraint of the form:
\begin{equation}
    \ell \leq g(\{x_i\}_{i \in V}) \leq u,
    \label{eq:constraint}
\end{equation}
where $g(\{x_i\}_{i \in V}) \in \mathbb{Z}$ is a function of binary variables $x_i \in \{0, 1\}$ and $\ell, u \in \mathbb{Z}$\footnote{For general constraints, we define $\hat{H}_\mathrm{cs}=\sum_{i\in V} \epsilon_i \hat{\sigma}_i^z$ for feasible solutions and $\hat{H}_\mathrm{cs}=1+\sum_{i\in V} \epsilon_i \hat{\sigma}_i^z$ for infeasible solutions, where $\epsilon_i$ are small random variables. However, in this work, we focus on constraints of the form given in~\eqref{eq:constraint}.}.
The inequality constraint includes an equality constraint when $\ell = u$.
Then a compressed-space Hamiltonian is given by
\begin{equation}
    \hat{H}_\mathrm{cs} = \left( g(\{\frac{\hat{\sigma}_i^z+1}{2}\}) - \ell \right) \left( g(\{\frac{\hat{\sigma}_i^z+1}{2}\}) - u \right) + \sum_{i \in V} \epsilon_i \hat{\sigma}_i^z,
    \label{eq:csHamiltonian}
\end{equation}
where $\epsilon_i$ is a small random variable\footnote{For a single inequality constraint, $g(\{x_i\}) \geq \ell$ or $g(\{x_i\}) \leq u$, the compressed-space Hamiltonian is given by $\hat{H}_\mathrm{cs}=- (g(\{\frac{\hat{\sigma}_i^z+1}{2}\}) -\ell) + \sum_{i \in V} \epsilon_i \hat{\sigma}_i^z$ or $\hat{H}_\mathrm{cs}= g(\{\frac{\hat{\sigma}_i^z+1}{2}\}) - u + \sum_{i \in V} \epsilon_i \hat{\sigma}_i^z$, respectively.}.
Now we show that this Hamiltonian satisfies the conditions (i)-(iv).
Condition~(i) is trivially satisfied.
Condition~(ii) is satisfied since $\hat{H}_\mathrm{cs}$ consists of only $\hat{\sigma}_i^z$ operators.
To show condition~(iii), let us denote the largest energy of feasible solutions by $E_\mathrm{F}^\mathrm{(max)}$ and the smallest energy of infeasible solutions by $E_\mathrm{IF}^\mathrm{(min)}$.
$E_\mathrm{F}^\mathrm{(max)}$ is bounded from above as $E_\mathrm{F}^\mathrm{(max)} \leq N \max_{i \in V} |\epsilon_i|$ and $E_\mathrm{IF}^\mathrm{(min)}$ is bounded from below as $E_\mathrm{IF}^\mathrm{(min)} \geq 1-N \max_{i \in V} |\epsilon_i|$.
Therefore, if $|\epsilon_i|$ is sufficiently small such that $\max_{i \in V} |\epsilon_i| < 1/(2N)$, then condition~(iii) holds (i.e., $E_\mathrm{F}^\mathrm{(max)} < E_\mathrm{IF}^\mathrm{(min)}$).
%\begin{equation}
%E_\mathrm{F}^\mathrm{(max)} \leq N \max_{i\in V} |\epsilon_i| < 1-N \max_{i \in V} |\epsilon_i| \leq E_\mathrm{IF}^\mathrm{(min)}.
%\end{equation}
Condition~(iv) is ensured by the second term in~(\ref{eq:csHamiltonian}).

$\hat{U}_\mathrm{cs}$ is obtained by solving a minimization problem that finds a unitary operator $\hat{U}: \mathcal{H} \to \mathcal{H}$ to minimize
\begin{equation}
    E(\hat{U}, \hat{H}_\mathrm{cs})=2^{-m} \sum_{q\in \{0,1\}^m} \bra{0q} \hat{U} \hat{H}_\mathrm{cs} \hat{U}^\dagger \ket{0q},
    \label{eq:EU}
\end{equation}
where $\ket{0q} \coloneqq \ket{0^{N-m}}\ket{q}$.
%$E(\hat{U}, \hat{H}_\mathrm{cs}) = \mathrm{Tr} [\hat{H}_\mathrm{cs} \rho_U]$ where $\rho_U=\hat{U} (\ket{0^{N-m}} \bra{0^{N-m}} \otimes \hat{I}^m/2^m) U^\dagger$.
The optimized unitary operator $\hat{U}_* \coloneqq \argmin_{\hat{U}} E(\hat{U}, \hat{H}_\mathrm{cs})$ satisfies~(\ref{eq:compress}), as shown below.
$\hat{U}_*^\dagger$ maps the compressed space $\{\ket{0q}\}_{q\in \{0,1\}^m}$ to a subspace spanned by $\{\ket{\psi_i}\}_{i=1}^{2^m}$, where $\ket{\psi_i}$ is the $i$-the eigenstate of $\hat{H}_\mathrm{cs}$, with eigenenergies $\{E_i\}_{i=1}^{2^N}$ arranged in the ascending order, i.e., $E_1 < E_2< \ldots < E_{2^N}$.
The low-energy subspace is herein denoted by $\mathcal{H}_\mathrm{L}\coloneqq \mathrm{Span} (\{\ket{\psi_i}\}_{i=1}^{2^m})$.
Condition~(iii) ensures that $\mathcal{H}_\mathrm{F} 
\subseteq \mathcal{H}_\mathrm{L}$.
Therefore, for any $\ket{\psi} \in \mathcal{H}_\mathrm{F}$, there exists $\ket{\phi} \in \mathcal{H}_\mathrm{cs}$ such that $\hat{U}_* \ket{\psi} = \ket{0^{N-m}}\ket{\phi}$.
%Namely, for any ,
%\begin{equation}
%\ket{\psi} = \hat{U}_*^\dagger \ket{0^{N-m}} \ket{\phi} \in \mathrm{Span} ( \{ \ket{\psi_i}\}_{i=1}^{2^m} ) \supset \mathcal{H}_\mathrm{F},
%\end{equation}
%where $\ket{\psi_i}$ is the $i$-the eigenstate of $\hat{H}_\mathrm{cs}$, with eigenenergies $\{E_i\}_{i=1}^{2^N}$ arranged in the ascending order, i.e., $E_1 < E_2< \ldots < E_{2^N}$.

Conditions (ii) and (iv) guarantee that the mid-circuit measurements in Fig.~\ref{fig:qaoa}~(b) always yield $0$ states (i.e., the computation is not discarded) when $\hat{U}_*$ is implemented in the quantum circuit.
Let $p_\mathrm{dis}^{(j)}$ denote the discarding rate in the $j$-th layer of the quantum circuit, specifically defined as
\begin{equation}
    %p_\mathrm{dis}^{(i)}=1- \mathrm{Tr} |\langle \psi^{(i)} \ket{0^{N-m}}|^2,
    p_\mathrm{dis}^{(j)}=1- \| \langle 0^{N-m} \ket{\psi^{(j)}} \|^2,
\end{equation}
where $\ket{\psi^{(j)}} \in \mathcal{H}$ is the quantum state when the measurement on the $j$-th layer is performed and $\| \ket{\phi} \| =\sqrt{\langle \phi \ket{\phi}}$.
In the optimized circuit, $\ket{\psi^{(j)}} = \hat{U}_* \hat{U}_C(\gamma) \hat{U}_*^\dagger \ket{0^{N-m}} \ket{\phi}$, with $\ket{\phi} \in \mathcal{H}_\mathrm{cs}$.
First, $\hat{U}_*^\dagger \ket{0^{N-m}} \ket{\phi} \in \mathcal{H}_\mathrm{L}$.
Then, $\hat{U}_C(\gamma)$ keeps the state within $\mathcal{H}_\mathrm{L}$, meaning $\hat{U}_C(\gamma) \hat{U}_*^\dagger \ket{0^{N-m}} \ket{\phi} \in \mathcal{H}_\mathrm{L}$, as conditions~(ii) and~(iv) indicate that each $\ket{\psi_i}$ is an eigenstate of $\hat{U}_C(\gamma)$.
Finally, applying $\hat{U}_*$ results in $\ket{\psi^{(j)}} = \ket{0^{N-m}} \ket{\phi'}$ with $\ket{\phi'} \in \mathcal{H}_\mathrm{cs}$.
Thus, $p_\mathrm{dis}^{(j)}=0$.
%\begin{equation}
%    \hat{U}_* \hat{U}_C(\gamma) \hat{U}_*^\dagger \ket{0^{N-m}} \ket{\phi} = \ket{0^{N-m}} \ket{\phi'}.
%\end{equation}

We consider two methods to calculate $E(\hat{U}, \hat{H}_\mathrm{cs})$.
The first method, based on~\cite{nakanishi2019subspace}, involves applying $\hat{U}^\dagger$ to an initial state $\ket{0q}$, followed by calculating the expectation value of $\hat{H}_\mathrm{cs}$, which yields $\bra{0q}\hat{U} \hat{H}_\mathrm{cs} \hat{U}^\dagger \ket{0q}$ (see Fig.~\ref{qc:scalable}~(a)).
By summing over different initial states, we obtain $E(\hat{U}, \hat{H}_\mathrm{cs})$.
This approach uses $N$~qubits, and its computational cost grows exponentially with $m$.

%\begin{equation}
%    E(\hat{U}, \hat{H}_\mathrm{cs})=2^{-m} \sum_{q\in \{0,1\}^m} \bra{0^{N-m}}\bra{q} \hat{H}_\mathrm{cs} \ket{0^{N-m}}\ket{q}
%\end{equation}

The second method, based on~\cite{wang2021variational,consigli2024variational}, is a scalable approach with respect to $m$, and uses $N+m$ qubits.
%Suppose the compressed space Hamiltonian is expressed as a linear combination of operator $\hat{O}_\alpha$, where $\hat{O}_\alpha$ represents the $\hat{\sigma}_i^z$ operators and their tensor products.
%In this case,
%\begin{equation}
%    E(\hat{U}, \hat{H}_\mathrm{cs}) = \sum_\alpha c_\alpha E(\hat{U}, \hat{O}_\alpha),
    \label{eq:sum}
%\end{equation}
%where $c_\alpha \in \mathbb{R}$ is the coefficient.
$E(\hat{U}, \hat{H}_\mathrm{cs})$ is determined by the quantum circuit shown in Fig.~\ref{qc:scalable}~(b), and is related to the probability of measuring the lower $N$-qubits, denoted by $P(\sigma,\hat{U})$ (the equivalent between~\eqref{eq:EU} and~\eqref{eq:E_UO} is given in Appendix~A.):
\begin{equation}
    %p_0=\frac{1+E(\hat{U}, \hat{O}_\alpha)}{2}.
    E(\hat{U}, \hat{H}_\mathrm{cs})=\sum_{\sigma} \bra{\sigma} \hat{H}_\mathrm{cs}\ket{\sigma} P(\sigma,\hat{U}).
    \label{eq:E_UO}
\end{equation}
%Summing over different $\hat{O}_\alpha$ yields $E(\hat{U}, \hat{H}_\mathrm{cs})$.
%The computational cost is proportional to the number of terms in \eqref{eq:sum}, which is polynomial in $N$ for most practically considered constraints.
%The quantum circuit consists of two main parts.
%The first part performs a Hilbert-Schmidt decomposition of the density matrix, whereas the second part carries out a computation similar to a swap test.

\begin{figure}
\begin{minipage}[]{0.45\linewidth}
%\centering
\subcaption{}
\scalebox{0.8}{
\begin{quantikz}
    \lstick[wires = 1]{$\ket{0^{N-m}}$} & \gate[wires=3]{\hat{U}^\dagger} &\\
    \lstick[wires = 2]{$\ket{q}$}  & &\\
    {}  &  &
\end{quantikz}
}
\end{minipage}\hspace{-1cm}
\begin{minipage}[]{0.45\linewidth}
%\centering
\subcaption{}
\scalebox{0.6}{
\begin{quantikz}[wire types={q,q,q,q,q}]
%    &&&&&&\ket{0} &\setwiretype{q}&\gate{H}&\ctrl{1}&\ctrl{2}&\ctrl{4}&\gate{H}&\meter{}&\\
%    \lstick[wires = 1]{$\ket{0^{N-m}}$} & {} &{}&{}&\gate[wires=3]{\hat{U}^\dagger}&&&&\gate[wires=2]{\hat{O}_\alpha}&\swap{3}&{}&{}&{}&{}&\\
    \lstick[wires = 2]{$\ket{0^{m}}$} & \gate[wires=1]{H} &\ctrl{3}&{}&{}&\\
    {} & \gate[wires=1]{H} &{}&\ctrl{3}&{}&\\
    \lstick[wires = 3]{$\ket{0^{N}}$} & {} &{}&{}&\gate[wires=3]{\hat{U}^\dagger}&\meter{}&\setwiretype{n}\rstick[wires = 3]{$P(\sigma;\hat{U})$}\\
     & {} &\targ{}&{}&{}&\meter{}&\setwiretype{n}\\
    {} & {} &{}&\targ{}&{}&\meter{}&\setwiretype{n}
\end{quantikz}
}
\end{minipage}
\caption{Two different quantum circuits to compute $E(\hat{U},\hat{H}_\mathrm{cs})$.
(a) The approach uses $N$ qubits and $2^m$ different initial states $\ket{0^{N-m}} \otimes \ket{q}$, where $q \in \{0, 1\}^m$.
(b) The approach uses $N+m$ qubits.
$P(\sigma;\hat{U})$ is the probability distribution function of the measurement on the lower $N$~qubits.
%The circuit calculates $E(\hat{U},\hat{O}_\alpha)$, where $\hat{O}_\alpha$ acts on $2$~qubits.
%The circuit configurations shown before and after the dotted vertical line represent the Hilbert-Schmidt decomposition of the density matrix (on the left) and a computation analogous to a swap test (on the right), respectively.
}
\label{qc:scalable}
\end{figure}

We use a variational approach to solve the minimization problem $\min_{\hat{U}} E(\hat{U}, \hat{H}_\mathrm{cs})$ using two types of circuit ansatz.
The first, referred to as continuous-parametric ansatz (C-ansatz), consists of $\ell$ layers of Y-rotations and controlled Y-rotations as depicted in Fig.~\ref{fig:compress_ansatz}~(a).
This type of ansatz was used in~\cite{yoshioka2020variational}.
Each Y-rotation gate, labeled $i$, has a real-valued parameter $\theta_i$.
It is represented as $\hat{R}_y^{(j)} = \exp (-i \theta_i \hat{\sigma}_j^y)$.
The second ansatz, called discrete-parametric ansatz (D-ansatz), consists of $\ell$ layers of X-gates, controlled X-gates (CNOT-gates), and controlled swap gates (CSWAP gates).
The choice of these gates is inspired by $\hat{U}_{\mathrm{binary},V}$ in Fig.~\ref{fig:deterministic}~(a).
%We call this ansatz classical because it does not generate entanglement from a product state.
The D-ansatz assigns a binary variable $0$ or $1$ to each gate and uses only the gates assigned a value of $1$.
For example, in Fig.~\ref{fig:compress_ansatz}~(b), the gates with a value of $1$ are highlighted in red\footnote{Note that each circuit layer can have a different combination of gates.}.
The variational parameters of the C-ansatz and D-ansatz are determined by a classical optimizer to minimize $E(\hat{U}, \hat{H}_\mathrm{cs})$.

\begin{figure}
\begin{minipage}[]{0.45\linewidth}
%\centering
\subcaption{C-ansatz}
\scalebox{0.6}{
\begin{quantikz}
    \lstick[wires = 1]{$\ket{0^{N-m}}$} &\gate[wires=1]{\hat{R}_y}&\ctrl{1} \gategroup[3,steps=7,style={inner sep=6pt}]{$\times \ell$}&\ctrl{2}&\gate[wires=1]{\hat{R}_y}&{}&\gate{\hat{R}_y}&{}&\gate{\hat{R}_y}&\rstick[wires = 3]{$\hat{U}^\dagger \ket{0^{N-m}} \ket{\phi}$}\\
    \lstick[wires = 2]{$\ket{\phi}$}  &{}&\gate[wires=1]{\hat{R}_y}&&\ctrl{-1}&\ctrl{1}&{}&\gate{\hat{R}_y}&\gate[wires=1]{\hat{R}_y}&{}\\
    {}  &&{}&\gate[wires=1]{\hat{R}_y}&&\gate[wires=1]{\hat{R}_y}&\ctrl{-2}&\ctrl{-1}&\gate[wires=1]{\hat{R}_y}&{}
\end{quantikz}
}
\end{minipage}\\
\begin{minipage}[]{0.45\linewidth}
%\centering
\subcaption{D-ansatz}
\scalebox{0.6}{
\begin{quantikz}
    \lstick[wires = 1]{$\ket{0^{N-m}}$} &\targ{} &\ctrl[style=red]{1} \gategroup[3,steps=10,style={inner sep=6pt}]{$\times \ell$}&\swap{1}&\swap{1}&\ctrl{1}&\ctrl[style=red]{2}&\targ[style=red]{}&{}&\targ[style=red]{}&{}&\targ{}&\rstick[wires = 3]{$\hat{U}^\dagger \ket{0^{N-m}} \ket{\phi}$}\\
    \lstick[wires = 2]{$\ket{\phi}$}  &{}&\swap[style=red]{1}& \ctrl{1} &\targX{}&\targ{}&{}&\ctrl[style=red]{-1}&\ctrl{1}&{}&\targ{}&\targ[style=red]{}&{}\\
    {}  & {} &\targX[style=red]{}&\targX{}&\ctrl{-1}&{}&\targ[style=red]{}&{}&\targ{}&\ctrl[style=red]{-2}&\ctrl{-1}&\targ{}&{}
\end{quantikz}
}
\end{minipage}
\caption{Two types of variational quantum-circuit ansatz adopted to engineer a compressed space.
$\ell$ denotes the number of the circuit layers.}
\label{fig:compress_ansatz}
\end{figure}

\subsubsection{Multiple constraint case}
\begin{algorithm}[t]
\caption{Engineering compressed space for a COP with multiple constraints} \label{Algo:engineer}
\begin{algorithmic}[1]
%% Input / Output
\renewcommand{\algorithmicrequire}{\textbf{Input:}}
%¥Require $¥sigma'$, $M$, and $V^m$ for $m ¥in ¥{1, ¥ldots, M¥}$
\Require $g_k(\{x_i\}_{i\in V})\in [\ell_k, u_k]$ for $k \in M$ (see \eqref{eq:feasible})
\renewcommand{\algorithmicensure}{\textbf{Output:}}
\Ensure $\hat{U}_\mathrm{cs}$
%% Body
\State $\hat{U}^{(0)} \gets \hat{I}^{N}$
\State $m^{(0)} \gets N$
\For{$k=0$ to $|M|-1$}
    \State Update $m^{(k)}$ to $m^{(k+1)} (\leq m^{(k)})$
    \State Generate $\hat{U}'$ for the $k+1$th constraint %using a deterministic method or a variational method
    \State $\hat{U}^{(k+1)} \gets \hat{U}' \hat{U}^{(k)}$
\EndFor
\State $\hat{U}_\mathrm{cs} \gets \hat{U}^{(|M|)}$
%¥For{$m = 1$ to $M$}
%        \State $f \gets \mathsf{True}$
%        \While{$f = \mathsf{True}$}
%                \For{$i \in V$}
                        %¥State $¥Delta Q_i ¥gets$ Energy change of $Q'$ ¥par
                        %hskip¥algorithmicindent ¥hspace{15mm} when flipping $x_i$
%                        \State $\Delta Q_i \gets$ Energy change of $Q'$ when flipping $x_i$
%                \EndFor
%                \State $j \gets \min(\argmin_{i \in V} \Delta Q_i$)
%                \If{$\Delta Q_j < 0$}
%                        \State $x_j \gets 1-x_j$
%                \Else
%                        \State $f \gets \mathsf{False}$
%                \EndIf
%        \EndWhile
%¥EndFor
%\State $\sigma_i \gets 2 x_i - 1$ for $i \in V$
\end{algorithmic}
\end{algorithm}

Algorithm~\ref{Algo:engineer} provides a flow to generate $\hat{U}_\mathrm{cs}$ for COPs with multiple constraints in \eqref{eq:feasible}.
The unitary operator begins with the identity operator, $\hat{U}^{(0)}=\hat{I}^{N}$ (line~1), and is sequentially updated by incorporating the constraints one by one.
Here, $\hat{U}^{(k)}$ represents a unitary operator that maps the feasible solution space for up to $k$ constraints to a compressed space with $m^{(k)}$ qubits.
To generate $\hat{U}^{(k+1)}$ from $\hat{U}^{(k)}$, the dimension of the compressed space is decreased (line~4) and generate $\hat{U}^{(k+1)}=\hat{U}' \hat{U}^{(k)}$ using a deterministic method or a variational method (line~5).
Sec.~\ref{subsec:qap} demonstrates a deterministic method to generate $\hat{U}'$.
Variational methods assign variational parameters for $\hat{U}'$ and determine them by minimizing the energy expectation value of the compressed-space Hamiltonian for the $(k+1)$-th constraint $\hat{H}_\mathrm{cs}^{(k)}$, while fixing $\hat{U}^{(k)}$, i.e., $\min_{\hat{U}'} E(\hat{U}' \hat{U}^{(k)}, \hat{H}_\mathrm{cs}^{(k)})$.
Iterating the processes gives $\hat{U}_\mathrm{cs}=\hat{U}^{(|M|)}$.

We provide a remark regarding the setting of $m^{(k+1)}$ (line~4).
Although $m^{(k+1)}$ is determined by the dimension of the feasible solution space, estimating this dimension is generally a challenging task.
One approach is to sample a solution from $\hat{U}^{(k)} \ket{0^{N-m^{(k)}}} \ket{\phi}$, where $\ket{\phi}$ is a random input, and evaluate whether it satisfies the $(k+1)$-th constraint.
This can offer an approximate estimation of the dimension of the feasible solution space.

%Second remark is on the necessity to consider each constraint subsequently when the constraints are dependent with each other.
%Here we call $i$-th and $j$-th constraints are dependent when $V_i \cap V_j \neq \varnothing$.  
%Consider an example of two constraints: $\sum_a$

\section{Applications of compressed space QAOA}
\label{sec:case}
We consider the Max-$k$ cut problems for $k\in\{3,4\}$, the QAPs, and the QKPs as demonstrations of the CS-QAOA.
They are categorized into NP-hard problems~\cite{frieze1997improved, pisinger2007quadratic,sahni1976pcomplete}.

\subsection{Ising models for COPs}
%Ising model for COPs with constraints reads
%\begin{equation}
%    \hat{H}=\frac{1}{\mathcal{N}} (\hat{H}_\mathrm{obj}+A \hat{H}_\mathrm{cst}),
%\end{equation}
%where $\hat{H}_\mathrm{obj}$ and $\hat{H}_\mathrm{cst}$ are Ising models for the objective function and the constraints, respectively.
%$A$ is a constraint coefficient and $\mathcal{N}$ is a normalization constant so that the maximal absolute value of $\{J_{ij}\}_{(i,j)\in E}$ and $\{h_i\}_{i \in V}$ is one.
%Here, we provide the Ising models for the COPs considered.
Here, we present QUBOs (quadratic unconstrained binary optimizations)~\cite{tanahashi2019application}, which are equivalent to the Ising models, for the COPs to be studied.
The QUBO has a form of
\begin{equation}
    Q=\sum_{i=1}^N \sum_{j=1}^N Q_{ij} x_i x_j +Q_0,
    \label{eq:qubo}
\end{equation}
where $Q_{ij} \in \mathbb{R}$, $Q_0 \in \mathbb{R}$, and $x_i \in \{0, 1\}$.
The QUBO for COPs with constraints is given by
\begin{equation}
    Q=\frac{1}{\mathcal{N}} (Q_\mathrm{obj}+A Q_\mathrm{cst}),
    \label{eq:qubo2}
\end{equation}
where $Q_\mathrm{obj}$ and $Q_\mathrm{cst}$ represent the QUBOs for the objective function and the constraints, respectively.
$A$ is a constraint coefficient and $\mathcal{N}$ is a normalization constant.
The corresponding Ising model in~\eqref{eq:Ising} is obtained by the replacement of $x_i$ in \eqref{eq:qubo} by $(\hat{\sigma}_i^z + 1)/2$.
The normalization $\mathcal{N}$ is determined so that the maximal absolute value of $\{J_{ij}\}_{(i,j)\in E}$ and $\{h_i\}_{i \in V}$ of the Ising model is one.

The Max-$k$ cut problem partitions the vertex set of a graph $G_\mathrm{m}=(V_\mathrm{m}, E_\mathrm{m})$ into $k$~subsets such that the number of edges connecting the different subsets is maximized.
The binary variables $x_{i,s} \in \{0, 1\}$ are assigned for $i\in V_\mathrm{m}$ and $s \in \{1, \ldots, k\}$.
When $x_{i,s}=1$, vertex~$i$ belongs to the \mbox{$s$-th} subset.
Without loss of generality, we can fix vertex~$0$ to belong to the first subset.
The feasible solution space is given by one-hot constraints, $\sum_{s=1}^k x_{i,s}=1$ for $i \in V_\mathrm{m} \setminus \{0\}$.
This type of constraints also appears in problems like graph coloring and clique cover~\cite{lucas2014Ising}.
The QUBO for a \mbox{Max-$k$} cut problems is given as
\begin{align}
        Q_\mathrm{obj}&=-\sum_{\substack{(i,j) \in E_\mathrm{m}\\ i,j \in V_\mathrm{m} \setminus \{0\}}} \sum_{s=1}^k x_{i,s} x_{j,s}-\sum_{\substack{i \in V_\mathrm{m} \setminus \{0\}\\(i,0)\in E_\mathrm{m}}}x_{i,1},\nonumber\\
        Q_\mathrm{cst}&= \sum_{i \in V_\mathrm{m} \setminus \{0\}} \left( \sum_{s=1}^k x_{i,s} -1 \right)^2.
\end{align}
Total number of qubits for the Ising model is given by $N=k(|V_\mathrm{m}|-1)$.

The QAP assigns $n_\mathrm{f}$ facilities to the same number of locations so that the total cost is minimized given the flow amount $\{f_{ij} \}_{i,j=1}^{n_\mathrm{f}}$ between facilities and the distance $\{d_{ab}\}_{a,b=1}^{n_\mathrm{f}}$ between locations~\cite{koopmans1957assignment}.
The binary variables $x_{i,a} \in \{0, 1\}$ are assigned for $i \in \{1,\ldots,n_\mathrm{f}\}$ and $a \in \{1,\ldots,n_\mathrm{f}\}$.
When $x_{i,a}=1$, facility~$i$ is assigned to location~$a$.
The feasible solution space is given by one-hot constraints, $\forall a, \sum_{i=1}^{n_\mathrm{f}} x_{i,a}=1$ and $\forall i, \sum_{a=1}^{n_\mathrm{f}} x_{i,a}=1$.
This type of constraints also appears in the travel salesman problem~\cite{lucas2014Ising}.
The QUBO for a QAP is given as~\cite{lucas2014Ising,glover2018tutorial}
\begin{align}
    Q_\mathrm{obj}&=\sum_{i=1}^{n_\mathrm{f}} \sum_{\substack{j=1 \\(j\neq i)}}^{n_\mathrm{f}} \sum_{a=1}^{n_\mathrm{f}} \sum_{\substack{b=1 \\ (b\neq a)}}^{n_\mathrm{f}} f_{ij} d_{ab} x_{i,a} x_{j,b},\nonumber\\
    Q_\mathrm{cst}&=\sum_{i=1}^{n_\mathrm{f}} \left( \sum_{a=1}^{n_\mathrm{f}} x_{i,a} -1 \right)^2 + \sum_{a=1}^{n_\mathrm{f}} \left( \sum_{i=1}^{n_\mathrm{f}} x_{i,a} -1 \right)^2.
\end{align}
Total number of qubits for the Ising model is given by $N=n_\mathrm{f}^2$.

The QKP finds a set of items to yield the highest profit within the capacity of the knapsack.
The set of $n_\mathrm{i}$ items with weights $\{w_i\}_{i=1}^{n_\mathrm{i}}$ and profits $\{p_{ij}\}_{i\leq j}^{n_\mathrm{i}}$ and the knapsack capacity $C$ are given.
Here, $p_{ii}$ represents the profit of item $i$ and $p_{ij}$ for $i<j$ denotes the additional profit when items $i$ and $j$ are selected.
Binary variables $\{x_i\}_{i=1}^{n_\mathrm{i}}$ are assigned such that $x_i=1$ when item $i$ is selected and $x_i=0$ when item $i$ is not selected.
The feasible solution space is given by an inequality constraint, $\sum_{i=1}^{n_\mathrm{i}} w_i x_i \leq C$.
The QUBO for a QKP is given as~\cite{montanez2024unbalanced}
\begin{align}
        Q_\mathrm{obj}&=- \sum_{i=1}^{n_\mathrm{i}} \sum_{j=i}^{n_\mathrm{i}} p_{ij} x_i x_j,\nonumber\\
        Q_\mathrm{cst}&=\left(\sum_{i=1}^{n_\mathrm{i}} \frac{w_i x_i}{C}\right)^2.
\end{align}
Total number of qubits for the Ising model is given by $N=n_\mathrm{i}$.

\subsection{Instances of COPs}
We generate ten instances for the Max-$k$ cut problems and the QAPs and six instances for the QKPs for each problem size.
The Max-$k$ cut problem instances are specified by the random graph $G_\mathrm{m}=(V_\mathrm{m}, E_\mathrm{m})$, which is generated by connecting each pair of vertices with probability $0.5$.
The QAP instances are specified by the flow amount $f_{ij}$ and the distance $d_{ab}$, which are symmetric (i.e., $f_{ij}=f_{ji}$ and $d_{ab}=d_{ba}$).
Each of $f_{ij}$ and $d_{ab}$ is randomly and uniformly selected from the sets $\{1,\ldots, 5\}$ and $\{1,\ldots, 10\}$, respectively.
The QKP instances are created by using benchmarks with $100$ and $200$ items listed  in~\cite{patvardhan2015solving}.
They are labeled as $n\_\mathrm{100}\_i$, where $n \in \{100, 200\}$ is the number of items and $i$ denotes the problem instance.
Each instance has a profit matrix $\{p_{jk}^{(i)}\}_{j\leq k}^n$, weight $\{w_j^{(i)}\}_{j=1}^n$, and knapsack capacity $C_n^{(i)}$.
To generate the QKP instances with $n_\mathrm{i}$ items, we use a profit matrix $\{p_{jk}^{(i)}\}_{j\leq k}^{n_\mathrm{i}}$, weight $\{w_j^{(i)}\}_{j=1}^{n_\mathrm{i}}$, and knapsack capacity $\lfloor n_\mathrm{i} C_n^{(i)}/n \rfloor$.
Here, $\lfloor x \rfloor = \max \{m \in \mathbb{Z}| m \leq x\}$ is the floor function.
This work adopts problem instances where the feasible solution ratios $p_\mathrm{F}=|F|/2^N$ for the generated instances lie between $0.1$ and $0.5$ (see Table~\ref{Table:compress} for the value of $p_\mathrm{F}$).
The optimal solutions for each problem instance are determined using a brute-force search.

\subsection{Experimental condition}
%The tolerance ($ftol$) is chosen to be $0.001$.
%We used a variational approach to solve the minimization problem $\min_U E(U, H_\mathrm{cs})$ using two types of circuit ansatz for $U$.
The performances of the CS-QAOA and the conventional QAOA are compared using success probability averaged over the problem instances for each problem size.
The success probability is given as
\begin{equation}
    p_\mathrm{suc}= \left(\prod_{i=1}^p (1-p_\mathrm{dis}^{(i)}) \right) \sum_{x \in S}|\langle x \ket{\vec{\beta}^*,\vec{\gamma}^*}|^2,
\end{equation}
where $S$ is a set of optimal solutions for a COP.
%Here, $p_\mathrm{dis}^{(i)}$ denotes the discarding rate in the $i$-th layer of the optimized quantum circuit.
%Specifically,
%\begin{equation}
    %p_\mathrm{dis}^{(i)}=1- \mathrm{Tr} |\langle \psi^{(i)} \ket{0^{N-m}}|^2,
%    p_\mathrm{dis}^{(i)}=1- \| \langle 0^{N-m} \ket{\psi^{(i)}} \|^2,
%\end{equation}
%where $\ket{\psi^{(i)}} \in \mathcal{H}$ is the quantum state when the measurement on the $i$-th layer is performed and $\| \ket{\phi} \| =\sqrt{\langle \phi \ket{\phi}}$.
We obtain $\vec{\beta}^*$ and $\vec{\gamma}^*$ using the Powell optimizer implemented on SciPy with $11$ different initial sets of $(\vec{\beta}, \vec{\gamma})$.
One set uses $(\vec{\beta}, \vec{\gamma})=(\vec{0},\vec{0})$ according to~\cite{wiersema2020exploring, park2024hamiltonian} and the others select $\{ \beta_i \}_{i=1}^p$ and $\{ \gamma_i \}_{i=1}^p$ randomly taken from the interval $[0, 2\pi)$~\footnote{Although $\hat{H}_\mathrm{C}$ is not integer due to normalization in \eqref{eq:qubo2}, we verified that the results remain qualitatively the same even when the interval is extended to $[0,4\pi)$.}.
We set the option of the optimizer $ftol$ to $0.001$ and the others to default.
We search for the optimal value of the constraint coefficient $A$ in \eqref{eq:qubo2} with the precision threshold set to $1$ for the Max-$k$ cut problem and $10$ for the QAP and the QKP.

The QKP adopts the variational method to determine $\hat{U}_\mathrm{cs}$, whereas the \mbox{Max-$k$} cut problem and the QAP use the deterministic method.
To check the success of generating $\hat{U}_\mathrm{cs}$ in the variational method, we introduce a survival rate of feasible solutions in the compressed space:
\begin{equation}
    p_\mathrm{sur}= \frac{1}{|F|}\sum_{x \in F} \| \bra{0^{N-m}} \hat{U} \ket{x} \|^2.
\end{equation}
%where $F$ is a set of the feasible solutions.
The compressed space encompasses more feasible solutions as $p_\mathrm{sur}$ increases, and include all feasible solutions when $p_\mathrm{sur}=1$.
%The variational method uses the Powell optimizer implemented on SciPy for quantum ansatz (see Fig.~\ref{fig:compress_ansatz}~(a)), whereas simulated annealing for classical ansatz (see Fig.~\ref{fig:compress_ansatz}~(b)).

The C-ansatz (see Fig.~\ref{fig:compress_ansatz}~(a)) utilizes the Powell optimizer, with the same settings as described above, to optimize the real-valued parameters $\{\theta_i \}$, where the initial values are randomly selected from the interval $[0, \pi)$.
We use $n_\mathrm{rep}$ different initial sets of $\{\theta_i \}$, where $n_\mathrm{rep}$ is initially set to $10$, and search the optimal one that minimizes $E(\hat{U},\hat{H}_\mathrm{cs})$.
We set the threshold to $p_\mathrm{sur} \geq 0.98$.
If $p_\mathrm{sur} < 0.98$, we regenerate $\hat{U}$ by increasing $n_\mathrm{rep}$ and/or $\ell$.

The D-ansatz (see Fig.~\ref{fig:compress_ansatz}~(b)) uses a simulated annealing algorithm~\cite{kirkpatrick1983optimization} to optimize binary variables for each gate, where the initial values are randomly selected from $\{0, 1\}$.
We adopt a single-bit-flip Markov chain Monte Carlo method with the Metropolis algorithm\footnote{Here bit denotes the binary variable assigned to each gate.}.
We fix the number of inner loops to be the number of variational parameters, and set the number of outer loops $n_\mathrm{loop}$, initial temperature $T_\mathrm{i}$, and final temperature $T_\mathrm{f}$ as inputs.
We set the threshold to $p_\mathrm{sur} = 1$.
If $p_\mathrm{sur} < 1$, we regenerate $\hat{U}$ by increasing $n_\mathrm{loop}$ and/or $\ell$ or modifying $T_\mathrm{i}$ and $T_\mathrm{f}$.

We generate five $\hat{U}_\mathrm{cs}$ and obtain the median of the success probability over the five samples for each problem instance.
Then we calculate the average and the standard deviation of the success probability over the problem instances for each size.

All computations were conducted on a PC with 2.30GHz Inter(R) Xeon(R) \mbox{W-2195} CPU.
We used Python~3.9.2 integrated with Intel fortran as the implementation language.

\subsection{Coherent simulation results}
\subsubsection{Max-$k$ cut problem}
\begin{figure}
    \centering
    \includegraphics[width=0.49\linewidth]{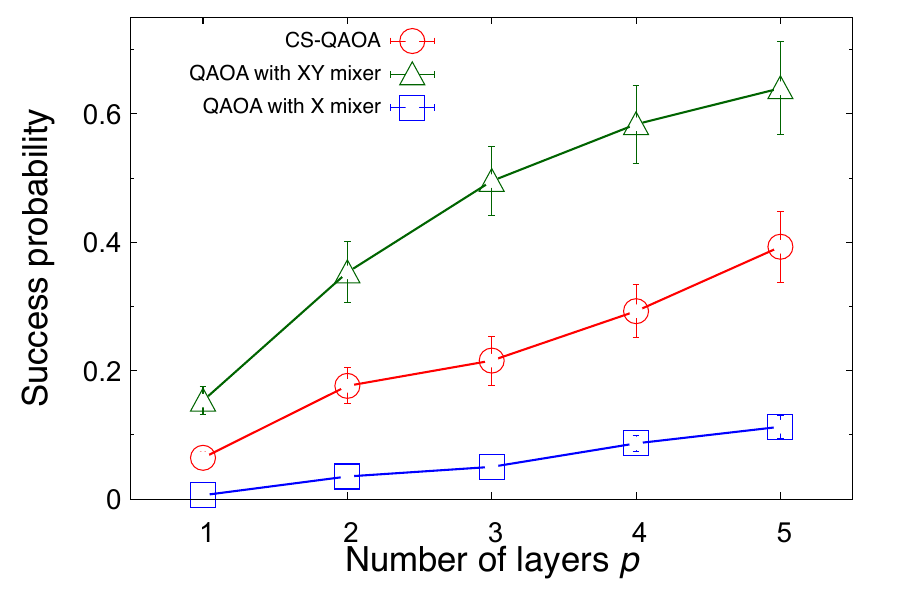}
    \includegraphics[width=0.49\linewidth]{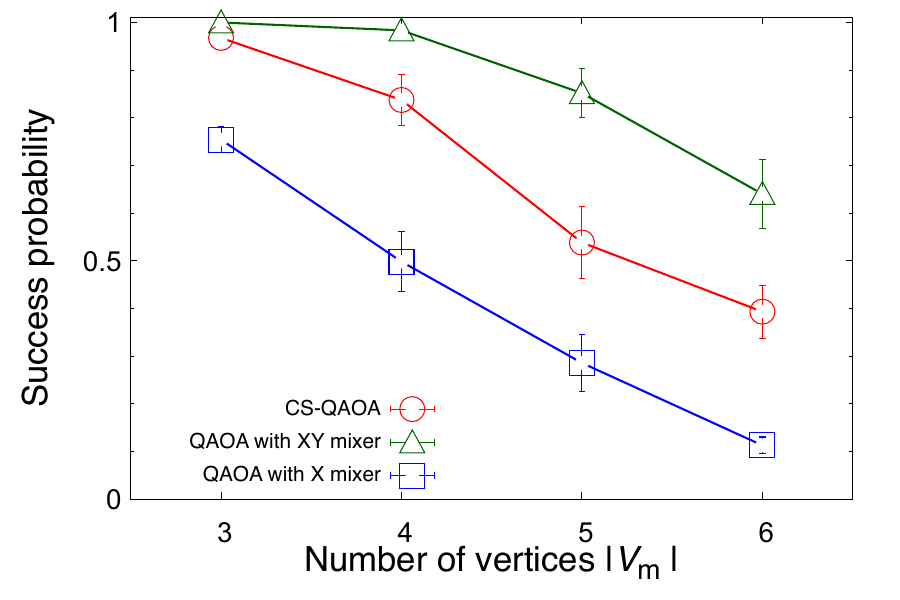}\\
    \includegraphics[width=0.49\linewidth]{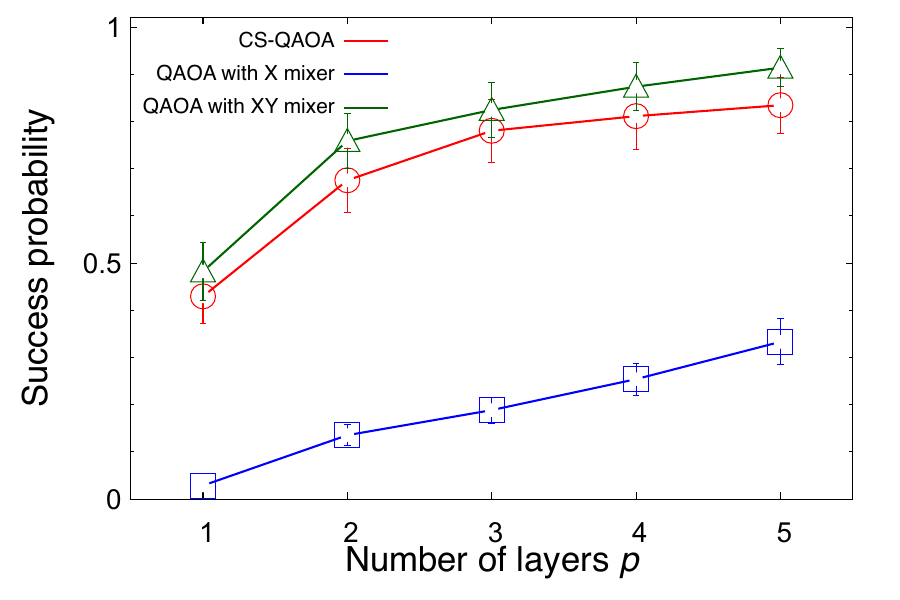}
    \includegraphics[width=0.49\linewidth]{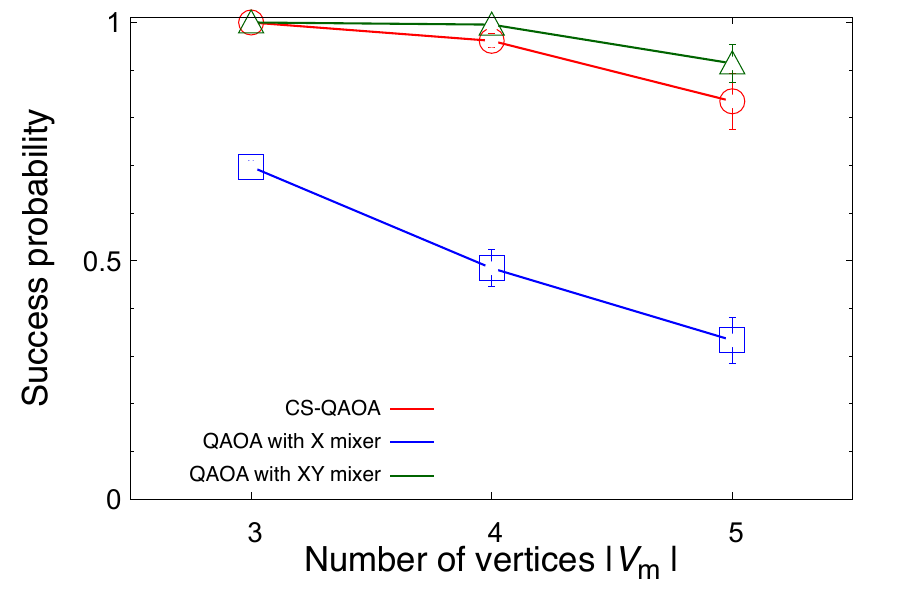}\\
    \includegraphics[width=0.49\linewidth]{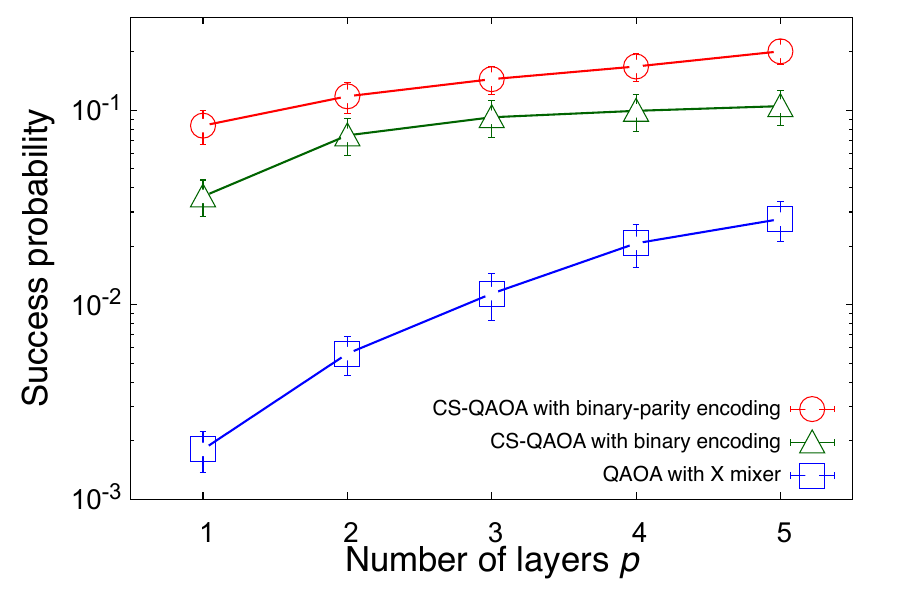}
    \includegraphics[width=0.49\linewidth]{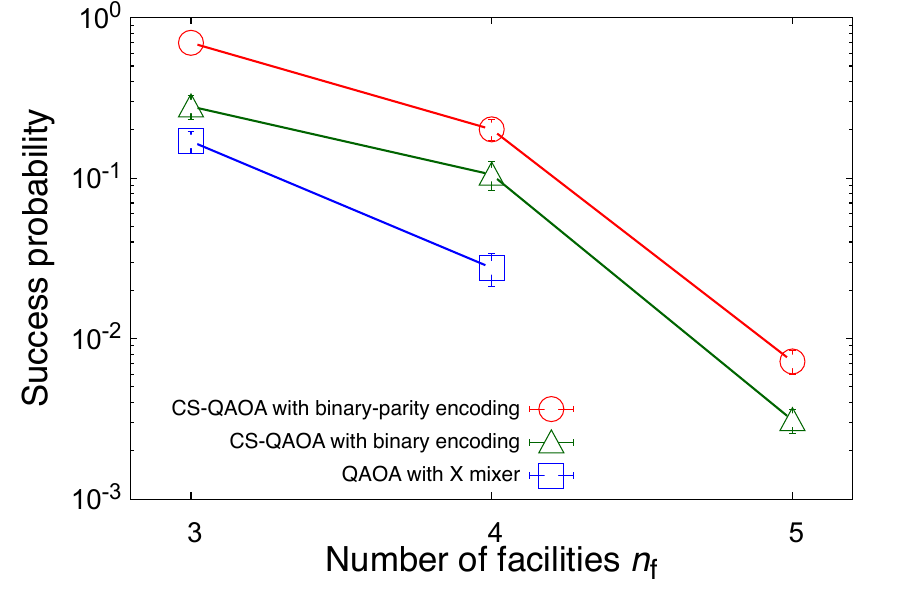}\\
    \includegraphics[width=0.49\linewidth]{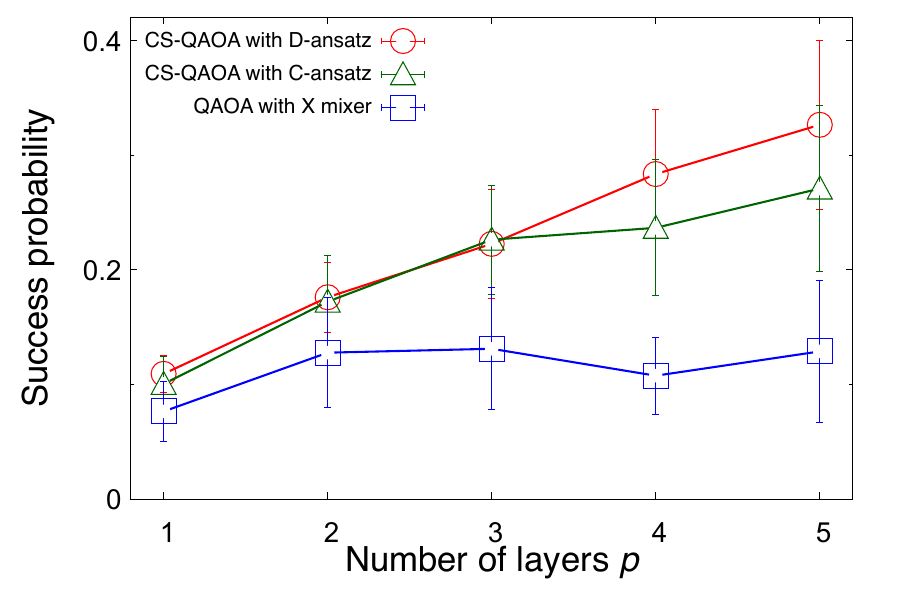}
    \includegraphics[width=0.49\linewidth]{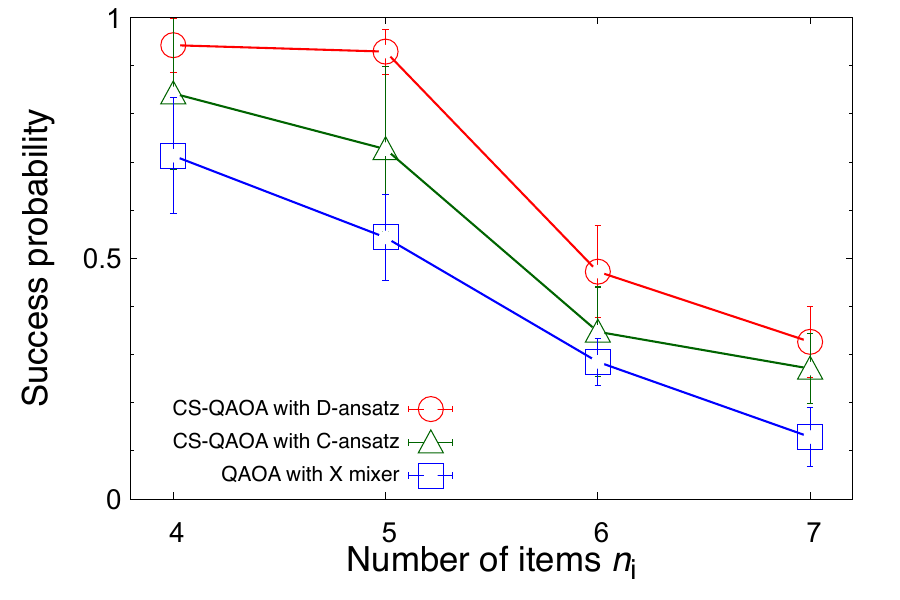}
    \caption{
    Dependences of the average success probability on the number of QAOA layers (left panels) and the problem sizes (right panels).
    Results for (top row) the Max-$3$ cut problems with $|V_\mathrm{m}|=6$ (left) and $p=5$ (right) and (second row) the Max-$4$ cut problems with $|V_\mathrm{m}|=5$ (left) and $p=5$ (right).
    %Success probability as a function of (left) the number of QAOA layers $p$ for $|V_\mathrm{m}|=6$ and (right) the number of vertices $|V_\mathrm{m}|$ at $p=5$.
    Circles, triangles, and squares denote the data for the CS-QAOA, the QAOA with the XY~mixer, and the QAOA with the X~mixer, respectively.
    (Third row) Results for the QAPs with $n_\mathrm{f}=4$ (left) and $p=5$ (right).
    %Success probability as a function of (left) the number of QAOA layers $p$ for $n_\mathrm{f}=4$ and (right) the number of facilities $n_\mathrm{f}$ at $p=5$.
    Circles, triangles, and squares denote the data for the CS-QAOA with binary-parity encoding, the CS-QAOA with the binary encoding, and the QAOA with the X~mixer, respectively.
    (Bottom row) Results for the QKPs with $n_\mathrm{i}=7$ (left) and $p=5$ (right).
    %Success probability as a function of (left) the number of QAOA layers $p$ for $n_\mathrm{i}=7$ and (right) the number of items $n_\mathrm{i}$ at $p=5$.
    Circles, triangles, and squares denote the data for the CS-QAOA with the D-ansatz, the CS-QAOA with the C-ansatz, and the QAOA with the X~mixer, respectively.
    The error bars in figures denote the standard deviation over the problem instances.
    }
    \label{fig:experiment}
\end{figure}

Since the Max-$k$ cut problem has the one-hot constraints for each vertex, $\hat{U}_\mathrm{cs}$ is given by
\begin{equation}
    \hat{U}_\mathrm{cs}=\prod_{i\in V_\mathrm{m}} \hat{U}_{\mathrm{binary}, V_i},
\end{equation}
where $V_i =\{(i,s)\}_{s=1}^k$.
%We consider Max-$k$ cut problems as a demonstration of compressed space QAOA with the deterministic method.
%This problem minimizes the number of cuts, where a cut denotes the edge connecting vertices with different colors given the graph $G(V, E)$ and $k$ colors.
%Here, binary variables $x_{ij}$ are assigned to denote the $j$th color at $i \in V$.
%Feasible solution space is given by one-hot constraints for each vertex, and thus the compressed space QAOA is applicable to this problem.

We compared the success probability among the CS-QAOA, the QAOA with the X~mixer, and the QAOA with the XY~mixer.
The initial state for the X~mixer is $\ket{\psi_\mathrm{ini}}=\ket{+^n}$, whereas for the XY~mixer, it is the product of the $W$-state for each vertex, i.e., $\ket{\psi_\mathrm{ini}}=\bigotimes_{i\in V_\mathrm{m} \setminus \{0\}} \ket{W_{V_i}}$.
Fig.~\ref{fig:experiment} (top and second rows) show the dependences of the success probability on $p$ and $|V_\mathrm{m}|$, respectively.
We found that the CS-QAOA outperforms the conventional QAOA with the X~mixer in both $k=3$ and $4$.
However, when $k=3$, the CS-QAOA performs worse  than the QAOA with the XY~mixer.
This degradation is attributed to the presence of some infeasible solutions in the compressed space.
In contrast, when $k=4$, where the compressed space completely excludes the infeasible solutions, the performance gap between the two methods becomes small.

\subsubsection{QAP}\label{subsec:qap}
%\begin{figure}
%    \centering
%    \includegraphics[width=0.49\linewidth]{./Figure/success_nloc4.pdf}
%    \includegraphics[width=0.49\linewidth]{./Figure/success_qap_p5.pdf}
%    \caption{
%    QAP result.
%    Success probability as a function of (upper) the number of QAOA layers $p$ for $n_\mathrm{f}=4$ and (lower) the number of facilities $n_\mathrm{f}$ at $p=5$.
%    Circles, triangles, and squares denote the data for CS-QAOA with binary-parity encoding, CS-QAOA with binary encoding, and QAOA with X~mixer, respectively.
%    }
%    \label{fig:QAP}
%\end{figure}

The feasible solution space for the QAPs is given by two types of one-hot constraints: $\forall a, \sum_{i=1}^{n_\mathrm{f}} x_{i,a}=1$ and $\forall i, \sum_{a=1}^{n_\mathrm{f}} x_{i,a}=1$.
When the first $n_\mathrm{f}$ constraints are considered, the unitary transformation for the compressed space is given by
\begin{equation}
    \hat{U}_1=\prod_{a=1}^{n_\mathrm{f}} \hat{U}_{\mathrm{binary}, V_a},
\end{equation}
where $V_a =\{(i,a)\}_{i=1}^{n_\mathrm{f}}$.
Each unitary transformation maps a one-hot vector $\{x_{n_\mathrm{f},a},\ldots,x_{1,a}\}$, i.e., a vector with a single nonzero (1) element and the rest being zeros, to a binary representation $\{x'_{n_\mathrm{f},a}, \ldots, x'_{1,a}\}$.
For example, a one-hot vector with $x_{4,a}=1$ when $n_\mathrm{f}=5$ gives the binary representation of $4-1$ as $\{x'_{i,a}\}=\{0,0,0,1,1\}$.
Note that the first two qubits are always in the zero state for one-hot vectors, since $\lceil \log_2 n_\mathrm{f} \rceil=3$.

Next, we consider the remaining constraints: $\forall i, \sum_{i=1}^{n_\mathrm{f}} x_{i,a}\\=1$.
These constraints imply that one-hot vectors $\{x_{i,a}\}$ have the nonzero value at different $i$ among $a \in \{1,\ldots,n_\mathrm{f}\}$.
In other words, $\{x'_{i,a}\}_{i=1}^{n_\mathrm{f}}$ gives a different value from $0$ to $n_\mathrm{f}-1$ in the binary representation for $a \in \{1,\ldots,n_\mathrm{f}\}$.
This observation indicates that $\sum_{a=1}^{n_\mathrm{f}} x'_{i,a}$ should be even or odd for $i\in \{1,\ldots, \lceil \log_2 n_\mathrm{f} \rceil\}$.
For example, when $n_\mathrm{f}=5$, the binary representations from $0$ to $4$ give $\{0,0,0\}$, $\{0,0,1\}$, $\{0,1,0\}$, $\{0,1,1\}$, and $\{1,0,0\}$, and thus $\sum_{a=1}^{n_\mathrm{f}} x'_{1,a}=2$ (even), $\sum_{a=1}^{n_\mathrm{f}} x'_{2,a}=2$ (even), and $\sum_{a=1}^{n_\mathrm{f}} x'_{3,a}=1$ (odd).
Thus the unitary transformation is generally given by
\begin{equation}
    \hat{U}_2 = \prod_{i=1}^{\lceil \log_2 n_\mathrm{f} \rceil} \hat{U}_{\mathrm{even (odd)}, V_i},
\end{equation}
where $V_i =\{(i,a)\}_{a=1}^{n_\mathrm{f}}$.
Note that the parity of the $i$-th unitary operator, whether it is set as even or odd, is determined by $n_\mathrm{f}$ based on the above argument.

The CS-QAOA uses two types of unitary transformation:
\begin{equation}
    \hat{U}_\mathrm{cs}=\left\{
    \begin{aligned}
    &\hat{U}_2 \hat{U}_1 \quad \text{(binary-parity encoding)}\\
    &\hat{U}_1 \quad \text{(binary encoding)}    
    \end{aligned}
    \right.
\end{equation}
We call each case binary-parity encoding and binary encoding, respectively.
The value of $m$ is $(n_\mathrm{f}-1) \lceil \log_2 n_\mathrm{f} \rceil$ for the binary-parity encoding, whereas $n_\mathrm{f} \lceil \log_2 n_\mathrm{f} \rceil$ for the binary encoding.

We compare the success probability among the CS-QAOA with the binary-party encoding, the CS-QAOA with the binary encoding, and the conventional QAOA with the X~mixer.
Fig.~\ref{fig:experiment} (third row) shows the dependences of the success probability on $p$ and $n_\mathrm{f}$, respectively.
We find that the CS-QAOA with the binary-parity encoding outperforms others.
This result indicates that the compressed space with fewer qubits can efficiently explore the (near-)optimal solutions, at least for this problem.

\subsubsection{QKP}
\begin{table*}[h]
  \centering 
  \caption{The parameter set to generate the compressed space with the variational method. 
  The medians of the five samples used to generate the unitary operators $\hat{U}_\mathrm{cs}$ are presented.} 
  \scalebox{1}{
  \begin{tabular}{ccc|cccc|cccccc}\hline
   {} & {} & {} & \multicolumn{4}{c|}{C-ansatz} & \multicolumn{6}{c}{D-ansatz} \\ \hline
   Instance & $N$ & $p_\mathrm{F}$ & $m$ & $\ell$ & $n_\mathrm{rep}$ & $p_\mathrm{sur}$ & $m$ & $\ell$ & $n_\mathrm{loop}$ & $T_\mathrm{i}$ & $T_\mathrm{f}$ & $p_\mathrm{sur}$\\ \hline
   \multirow{4}{*}{$100\_100\_1$} & 4 & 0.5 & 3 & 1 & 10 & 1.000 & 3 & 2 & 1000 & 10 & 0.1 & 1\\
    & 5 & 0.3125 & 4 & 1 & 10 & 1.000 & 4 & 1 & 1000 & 10 & 0.1 & 1 \\
    & 6 & 0.28125 & 5 & 1 & 10 & 1.000 & 5 & 1 & 1000 & 10 & 0.1 & 1 \\
    & 7 & 0.2578125 & 6 & 1 & 10 & 1.000 & 6 & 1 & 1000 & 10 & 0.1 & 1 \\ \hline
   \multirow{4}{*}{$100\_100\_4$} & 4 & 0.3125 & 3 & 1 & 10 & 1.000 & 3 & 1 & 1000 & 10 & 0.1 & 1\\
    & 5 & 0.25 & 3 & 2 & 100 & 0.999 & 3 & 1 & 1000 & 10 & 0.1 & 1 \\
    & 6 & 0.375 & 5 & 2 & 100 & 0.999 & 5 & 1 & 1000 & 10 & 0.1 & 1 \\
    & 7 & 0.4375 & 6 & 3 & 100 & 0.982 & 6 & 1 & 10000 & 1 & 0.1 & 1 \\ \hline
    \multirow{4}{*}{$100\_100\_6$}& 4 & 0.125 & 1 & 1 & 10 & 1.000 &  1 & 1 & 1000 & 10 & 0.1 & 1\\
    & 5 & 0.15625 & 3 & 1 & 10 & 1.000 & 3 & 1 & 1000 & 10 & 0.1 & 1 \\
    & 6 & 0.1875 & 4 & 1 & 10 & 1.000 & 4 & 1 & 1000 & 1 & 0.01 & 1 \\
    & 7 & 0.2109375 & 6 & 2 & 10 & 1.000 & 5 & 1 & 100000 & 10 & 0.1 & 1 \\ \hline
    \multirow{4}{*}{$100\_100\_8$}& 4 & 0.25 & 2 & 1 & 10  & 1.000 &  2 & 1 & 1000 & 10 & 0.1 & 1\\
    & 5 & 0.25 & 3 & 1 & 10 & 1.000 & 3 & 1 & 1000 & 10 & 0.1 & 1 \\
    & 6 & 0.265625 & 5 & 1 & 10 & 1.000 & 5 & 1 & 1000 & 10 & 0.1 & 1 \\
    & 7 & 0.25 & 5 & 3 & 100 & 0.995 & 5 & 1 & 75000 & 10 & 0.05 & 1 \\ \hline
    \multirow{4}{*}{$200\_100\_2$}& 4 & 0.125 & 1 & 1 & 10 & 1.000 &  1 & 1 & 1000 & 10 & 0.1 & 1\\
    & 5 & 0.125 & 2 & 1 & 10 & 1.000 & 2 & 1 & 1000 & 10 & 0.1 & 1 \\
    & 6 & 0.15625 & 4 & 2 & 10 & 1.000 & 4 & 1 & 1000 & 10 & 0.1 & 1 \\
    & 7 & 0.1875 & 5 & 2 & 10 & 0.999 & 5 & 1 & 1000 & 10 & 0.1 & 1 \\ \hline
    \multirow{4}{*}{$200\_100\_10$}& 4 & 0.1875 & 2 & 1  & 10 & 1.000 & 2 & 1 & 1000 & 10 & 0.1 & 1\\
    & 5 & 0.1875 & 3 & 3 & 100 & 0.998 & 3 & 1 & 1000 & 10 & 0.1 & 1 \\
    & 6 & 0.234375 & 4 & 2 & 200 & 0.991 & 4 & 1 & 10000 & 10 & 0.1 & 1 \\
    & 7 & 0.28125 & 6 & 2 & 10 & 0.999 & 6 & 1 & 1000 & 10 & 0.1 & 1 \\
   \hline
   \end{tabular}
  }
  \label{Table:compress}
\end{table*}
%\begin{figure}[t]
%    \centering
%    \includegraphics[width=0.49\linewidth]{./Figure/success_qkp_n7.pdf}
%    \includegraphics[width=0.49\linewidth]{./Figure/success_qkp_p5.pdf}
%    \caption{
%    QKP result.
%    Success probability as a function of (upper) the number of QAOA layers $p$ for $n_\mathrm{i}=7$ and (lower) the number of items $n_\mathrm{i}$ at $p=5$.
%    Circles, triangles, and squares denote the data for CS-QAOA with classical ansatz, CS-QAOA with quantum ansatz, and QAOA with X~mixer, respectively.
%    }
%    \label{fig:QKP}
%\end{figure}
We consider QKPs as a demonstration of the CS-QAOA with a variational method.
The compressed-space Hamiltonian reads
\begin{equation}
\hat{H}_\mathrm{cs} =\left(\sum_{i=1}^N w_i \frac{\hat{\sigma}_i^z + 1}{2}\right) - C + \sum_{i=1}^N \epsilon_i \hat{\sigma}_i^z,
\end{equation}
where $\epsilon_i$ is a random variable uniformly taken from $[-0.05, 0.05)$.

We use both the C-ansatz and D-ansatz to engineer the compressed space.
Table~\ref{Table:compress} gives the parameter sets used for problem instances up to $N=7$.
Based on the QAP results, we aim to find a compressed space with the smallest possible $m$ such that $2^m \geq |F|$.
The C-ansatz fails to identify such a compressed space for the problem instance $100\_100\_6$ with $N=7$, whereas the D-ansatz successfully finds such a space for all problem instances.
%Table~\ref{Table:compress} indicates that the required computational cost (i.e., $\ell$, $n_\mathrm{rep}$, and $n_\mathrm{loop}$) typically increases with $N$.
The superior performance of the D-ansatz over the C-ansatz is likely due to its smaller parameter space (i.e., discrete vs. continuous).

We compare the success probability of the CS-QAOA with the C-ansatz and D-ansatz to the QAOA with the X~mixer.
Fig.~\ref{fig:experiment} (bottom row) show the dependences of the success probability on $p$ and $n_\mathrm{i}$, respectively.
We observed that the CS-QAOA with the D-ansatz consistently outperforms the others, while the CS-QAOA with the C-ansatz degrades in performance.

To understand the performance differences between the C-ansatz and D-ansatz, we examine the sample-to-sample fluctuations.
Specifically, we compare the standard deviation of the success probability between the C-ansatz and D-ansatz, averaged over the $5$ generated compression unitary operators.
As shown in Table~\ref{Table:s1s2}, the standard deviation for the C-ansatz is much larger than that for the D-ansatz.
This substantial fluctuation arises from the probability distribution function $p(x)$, which is produced by the unitary operator $\hat{U}_\mathrm{cs}^\dagger$ acting on the uniform state within the compressed space, i.e., $p(x)= |\bra{x} \hat{U}_\mathrm{cs}^\dagger \ket{0^{N-m}} \ket{+^m}|^2$.
While the D-ansatz yields the uniform distribution over $\{ \{x\} | \ket{x} \in \mathcal{H}_\mathrm{L} \}$, the C-ansatz leads to a nonuniform distribution.
The large standard deviation in the success probability results in a deterioration of performance on average.

%In addition, we investigate the dependence of the energy variance, $\mathrm{Var}[E]$, on the choice of ansatz, where the energy denotes $\bra{\vec{\beta},\vec{\gamma}} \hat{H}_\mathrm{Ising} \ket{\vec{\beta},\vec{\gamma}}$.
%Table~\ref{Table:s1s2} reports the average values of $\mathrm{Var}[E]$ for $p=5$,
%computed from $10,000$ samples with parameters $\beta_i\in[0, 2\pi)$ and $\gamma_i\in[0, 2\pi)$ drawn uniformly at random.
%We observe no significant difference in $\mathrm{Var}[E]$ between the two ansatz, which implies that the performance difference originates from differences in the distributions $p(x)$, rather than from energy landscape.

%\begin{table}[t]
%  \centering 
%  \caption{Comparison of $\Delta p_\mathrm{suc}$ and $\mathrm{Var[E]}$ between the C-ansatz and the D-ansatz, averaged over the $6$ problem instances at $p=5$.
%  The parenthesis denote the standard deviation.
%   } 
%  \scalebox{1}{
%  \begin{tabular}{c|cc|cc}\hline
%    & \multicolumn{2}{c|}{C-ansatz} & \multicolumn{2}{c}{D-ansatz}\\ \hline
%   $n_\mathrm{i}$ & $\Delta p_\mathrm{suc}$  & $\mathrm{Var}[E] $ & $\Delta p_\mathrm{suc}$ & $\mathrm{Var}[E]$ \\ \hline \hline
%   $4$ & $0.29 (0.09)$ & $0.078 (0.027)$ & $0.02 (0.02)$ & $0.044 (0.021)$\\ 
%   $5$ & $0.19 (0.07)$ & $0.066 (0.006)$ & $0.02 (0.01)$ & $0.072 (0.008)$\\
%   $6$ & $0.22 (0.06)$ & $0.055 (0.009)$ & $0.06 (0.02)$ & $0.084 (0.024)$\\
%   $7$ & $0.07 (0.02)$ & $0.036 (0.003)$ & $0.03 (0.01)$ & $0.037 (0.003)$\\ \hline
%   \end{tabular}
%  }
%  \label{Table:s1s2}
%\end{table}

\begin{table}[t]
  \centering 
  \caption{Comparison of the standard deviation of $p_\mathrm{suc}$ between the C-ansatz and the D-ansatz, averaged over the $6$ problem instances at $p=5$.
  The parenthesis denote the standard deviation.
   } 
  \scalebox{1}{
  \begin{tabular}{c|c|c}\hline
%    & \multicolumn{2}{c|}{C-ansatz} & \multicolumn{2}{c}{D-ansatz}\\ \hline
   $n_\mathrm{i}$ & C-ansatz  & D-ansatz \\ \hline \hline
   $4$ & $0.29 (0.09)$ & $0.02 (0.02)$ \\ 
   $5$ & $0.19 (0.07)$ & $0.02 (0.01)$ \\
   $6$ & $0.22 (0.06)$ & $0.06 (0.02)$ \\
   $7$ & $0.07 (0.02)$ & $0.03 (0.01)$ \\ \hline
   \end{tabular}
  }
  \label{Table:s1s2}
\end{table}

\subsection{Discussion on scalability}
\begin{figure}
    \centering
    \includegraphics[width=0.49\linewidth]{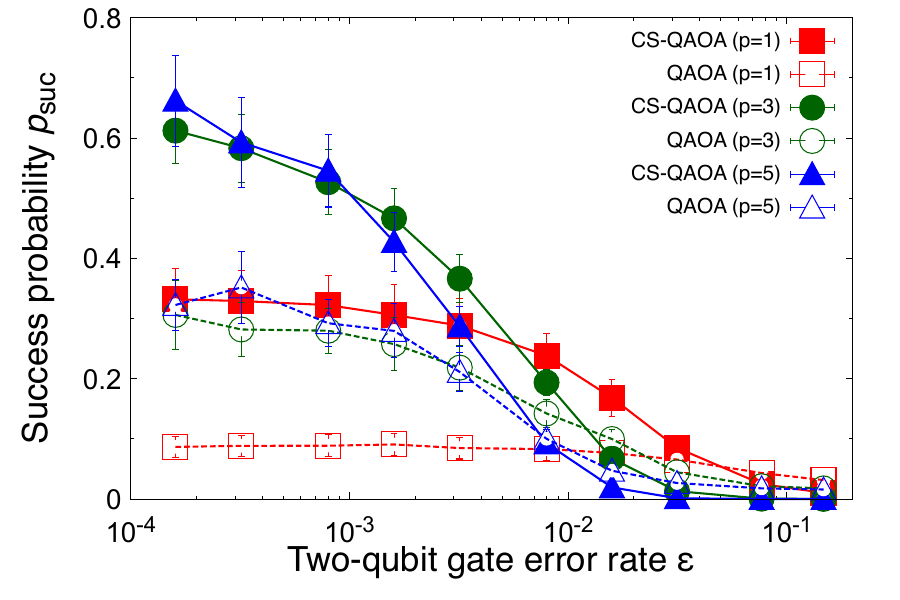}
    \includegraphics[width=0.49\linewidth]{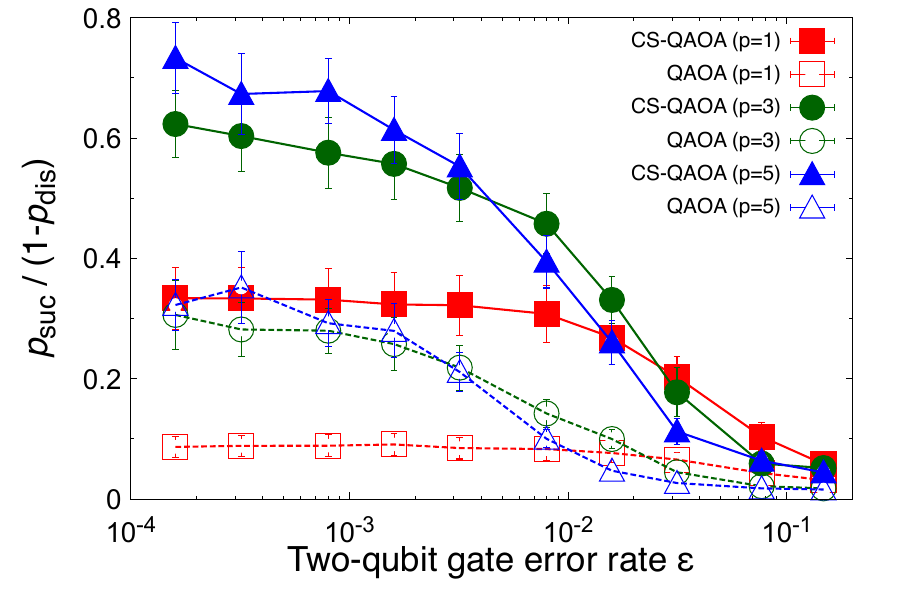}
    \caption{
    Noise simulation results for the Max-$3$ cut problems with $|V_\mathrm{m}|=4$.
    (Left) Total success probability $p_\mathrm{suc}$ and (Right) normalized success probability $p_\mathrm{suc}/(1-p_\mathrm{dis})$ as a function of the two-qubit gate error rate $\varepsilon$.
    }
    \label{fig:dissipation}
\end{figure}

Here, we discuss the scalability of the CS-QAOA from several perspectives.
First, we examine the impact of noise on its performance.
Specifically, we consider depolarizing noise that affects two-qubit gates, with the error rate denoted by $\varepsilon$ (see Appendix~B for details of the noise simulation).
The median two-qubit gate error rate in state-of-the-art processors is $\varepsilon\simeq 2\times 10^{-3}$~\cite{ibm}.
Fig.~\ref{fig:dissipation} (Left) presents the $\varepsilon$-dependences of $p_\mathrm{suc}$ for both the CS-QAOA and the QAOA across various values of $p$, using the Max-$3$ cut problems with $|V_\mathrm{m}|=4$.
The performance of both algorithms degrades with increasing $\varepsilon$.
The CS-QAOA exhibits a stronger sensitivity to noise than the QAOA.
For large $\varepsilon$, the QAOA outperforms the CS-QAOA, and the crossover point, where the QAOA becomes superior, shifts toward lower $\varepsilon$ as the number of layers increases.
A similar vulnerability has been reported in quantum annealing with the XY~mixer~\cite{igata2024constrained}.

However, the CS-QAOA allows for the mid-circuit measurement to detect and discard erroneous samples.
Fig.~\ref{fig:dissipation} (Right) shows the normalized success probability $p_\mathrm{suc}/(1-p_\mathrm{dis})$, where $p_\mathrm{dis}$ denotes the probability of discarding a sample:
\begin{equation}
    p_\mathrm{dis}=1-\prod_{i=1}^p (1-p_\mathrm{dis}^{(i)}).
\end{equation}
After this screening, the CS-QAOA consistently outperforms the QAOA across all values of $p$ and $\varepsilon$, indicating the benefit of intrinsic error detection.

To promote the CS-QAOA as a scalable method, it is essential to reduce $p_\mathrm{dis}$.
This rate is estimated as 
\begin{equation}
p_\mathrm{dis} \approx 1-\exp (-C\varepsilon p N_\mathrm{two}),
\end{equation}
where $C$ is a constant and $N_\mathrm{two}$ is the number of two-qubit gates per layer.
Since $N_\mathrm{two}$ typically increases with the problem size $N$, maintaining a high success probability becomes harder for larger instances under noise.
Addressing this scalability issue will require progress on both hardware and software fronts.
On the hardware side, reducing the gate error rate $\varepsilon$ is crucial.
On the software side, future work should focus on post-processing techniques or error correction methods that make use of the information obtained from mid-circuit measurements, even when errors are detected.

Second, we discuss the scalability of the variational method for constructing compression unitary operators in the CS-QAOA.
As shown in Table~\ref{Table:compress}, the computational cost (i.e., $\ell$, $n_\mathrm{rep}$, and $n_\mathrm{loop}$) generally increases with $N$.
This suggests that constructing $\hat{U}_\mathrm{cs}$ becomes increasingly difficult for constraints that involves a larger number of variables.
The C-ansatz and the D-ansatz contain $O(N^2 \ell)$ and $O(N^3 \ell)$ parameters, respectively.
The rapid growth in parameter count complicates the optimization landscape, which makes it harder to identify an optimal compressed space.
Moreover, it leads to an increase in $N_\mathrm{two}$, exacerbating the impact of noise.
To improve scalability, it is crucial to design more efficient ansatz that can construct compressed spaces with fewer parameters.
One possible direction is the use of regularization techniques to suppress parameter proliferation.

It is noted that the scalability issue can be avoided when considering a subset $M_0 \subseteq M$ in which the constraints are independent and each involves a small number of variables (e.g., constraints of the form $0\leq \sum_{i=1}^3 x_{3k+i} \leq 1$ for $k \in \{0,1,\ldots\}$).
Two constraints are called independent when $V_k \cap V_{k'} = \varnothing$ in \eqref{eq:feasible}.
In this case, $\hat{U}_\mathrm{cs}$ for the constraints in $M_0$ decomposes as
\begin{equation}
    \hat{U}_\mathrm{cs}=\bigotimes_{k\in M_0} \hat{U}_\mathrm{cs}^{(k)},
\end{equation}
where the compression unitary operator for constraint $k$, $\hat{U}_\mathrm{cs}^{(k)}$, acts non-trivially only on the qubits in $V_k$.
Importantly, once a suitable compression unitary operator is found via the variational method, it can be reused across different problem instances, as long as the underlying constraint remains unchanged.
This opens the possibility of building a constraint-specific database of $\hat{U}_\mathrm{cs}$, which could enhance the efficiency of this approach in practice (see Appendix~C for the list of $\hat{U}_\mathrm{cs}$ for typical inequalities of the form $\ell \leq \sum_i x_i \leq u$).

\begin{figure}
    \centering
    \includegraphics[width=0.6\linewidth]{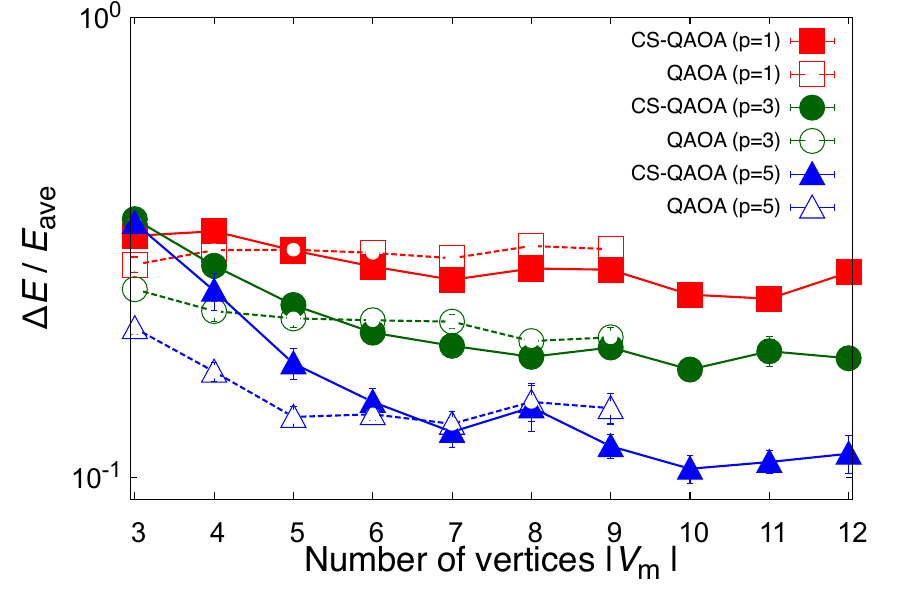}
    \caption{
    Normalized energy fluctuation $\Delta E/E_\mathrm{ave}$ as a function of $|V_\mathrm{m}|$ in the Max-$3$ cut problems.
    }
    \label{fig:barren}
\end{figure}
Finally, we examine the normalized energy fluctuation, $\Delta \bar{E}=\Delta E/E_\mathrm{ave}$, where $\Delta E$ and $E_\mathrm{ave}$ are the standard deviation and the average of the energy $\bra{\vec{\beta},\vec{\gamma}} \hat{H}_\mathrm{Ising} \ket{\vec{\beta},\vec{\gamma}}$ evaluated over randomly sampled parameters $\vec{\beta}$ and $\vec{\gamma}$, respectively.
Fig.~\ref{fig:barren} shows the dependence of $\Delta \bar{E}$ on $|V_\mathrm{m}|$ for the Max-$3$ cut problems at various values of $p$, using $100$~samples where each $\beta_i, \gamma_i\in[0, 2\pi)$ is drawn uniformly.
We assume an ideal (noise-free) quantum circuit.
The results indicate that $\Delta \bar{E}$ decreases with $|V_\mathrm{m}|$, but the decline is milder than the exponential, expected from the barren-plateau phenomenon~\cite{larocca2025barren}.
This behavior arises from the limited expressibility of quantum circuits with finite $p$.
In contrast, $\Delta \bar{E}$ exhibits an exponential decay with respect to $p$.
Thus, the algorithm is likely scalable as long as $p=O(1)$.
Moreover, we do not observe a significant difference between the CS-QAOA and the QAOA, especially for large $|V_\mathrm{m}|$, suggesting that restricting the search to a compressed space does not severely affect $\Delta \bar{E}$.
%In addition, we investigate the dependence of the energy variance, $\mathrm{Var}[E]$, on the choice of ansatz, where the energy denotes $\bra{\vec{\beta},\vec{\gamma}} \hat{H}_\mathrm{Ising} \ket{\vec{\beta},\vec{\gamma}}$.
%Table~\ref{Table:s1s2} reports the average values of $\mathrm{Var}[E]$ for $p=5$,
%computed from $10,000$ samples with parameters $\beta_i\in[0, 2\pi)$ and $\gamma_i\in[0, 2\pi)$ drawn uniformly at random.
%We observe no significant difference in $\mathrm{Var}[E]$ between the two ansatz, which implies that the performance difference originates from differences in the distributions $p(x)$, rather than from energy landscape.

\section{Conclusion}
\label{sec:conclusion}
We propose the CS-QAOA for solving constrained combinatorial optimization.
To extend the applicability of the CS-QAOA for various types of constraints, we develop methods for determining the unitary transformation between the compressed space and the original space on the gate-based quantum devices.
Coherent numerical simulations demonstrate its effectiveness over the Max-$3$ cut problems, the QAPs, and the QKPs.

In the future, we plan to implement the CS-QAOA on quantum hardware.
Our noisy simulations indicate that the validity of the CS-QAOA can be tested on current quantum devices, although its applicability is currently limited to small-scale instances.
The CS-QAOA can detect errors caused by noise through mid-circuit measurements, making it useful for mitigating errors on noisy quantum devices.
To broaden its practical use, further developments are required, including post-processing techniques for mid-circuit measurement outcomes and design of more efficient circuit ansatz.
The extension of the idea of compressed space engineering to other benchmarks in quantum optimization benchmarking library~\cite{koch2025quantum} and other algorithms such as variational quantum eigensolver and quantum machine learning is an important future direction.

\begin{appendices}
\section{Quantum circuit to evaluate $E(\hat{U},\hat{H}_\mathrm{cs})$}
This appendix provides a detailed explanation of the quantum circuit shown in Fig.~\ref{qc:scalable}~(b), which is used to evaluate $E(\hat{U},\hat{H}_\mathrm{cs})$.

%\begin{figure}
%\begin{minipage}[]{0.45\linewidth}
%\subcaption{}
%\scalebox{0.52}{
%\begin{quantikz}
%    \lstick[wires = 1]{$\ket{0^{N-m}}$} & &{}&{}&\gate[wires=3]{\hat{U}^\dagger}&{}&\\
%    \lstick[wires = 2]{$\ket{0^{m}}$} & \gate{H} &\ctrl{3}&{}&{}&{}&\\
%    {} & \gate{H} &{}&\ctrl{3}&{}&{}&\\
%    \lstick[wires = 3]{$\ket{0^{N}}$} & &{}&{}&\gate[wires=3]{\hat{U}^\dagger}&{}&\\
%     & &\targ{}&{}&{}&{}&\\
%    {} & &{}&\targ{}&{}&{}&
%\end{quantikz}
%}
%\end{minipage}
%\begin{minipage}[]{0.45\linewidth}
%\subcaption{}
%\scalebox{0.52}{
%\begin{quantikz}
%    \lstick[wires = 1]{$\ket{0}$} &\gate{H}&\ctrl{1}&\ctrl{2}&\ctrl{4}&\gate{H}&\rstick[wires = 7]{$U_O\ket{0}(\ket{\psi})^{\otimes 2}$}\\
%    \lstick[wires = 3]{$\ket{\psi}$} &\gate[wires=2]{\hat{O}_\alpha}&\swap{3}&{}&{}&{}&\\
%    {} &{}&{}&\swap{3}&{}&{}&\\
%    {} &{}&{}&{}&{}&{}&\\
%    \lstick[wires = 3]{$\ket{\psi}$} &{}&\targX{}&{}&\gate[wires=2]{\hat{O}_\alpha}&{}&\\
%    {} &{}&{}&\targX{}&{}&{}&\\
%    {} &{}&{}&{}&{}&{}&
%\end{quantikz}
%}
%\end{minipage}
%\caption{Decomposition of the quantum circuit in Fig.~\ref{qc:scalable}~(b) into (a) and (b).}
%\label{qc:scalable1}
%\end{figure}
Let us label the qubits in the circuit as $\{1,\ldots,N+m\}$ from top to bottom and analyze the output state of the circuit for the input state $\ket{0^{N+m}}$ by applying each gate sequentially.
The Hadamard gates $H$ applied to the qubits labeled from $1$ to $m$ give
\begin{equation}
    (H \ket{0})^m \ket{0^{N}}= \ket{+^m} \ket{0^{N}}.
\end{equation}
Then CNOT-gates create entanglement between the former $m$ qubits and the latter $m$ qubits:
\begin{align}
    \prod_{i=1}^m \mathrm{CX}_{i, N+i} \ket{+^m} \ket{0^{N}}= 2^{-\frac{m}{2}} \sum_{q \in \{0,1\}^m} \ket{q}\ket{0q}.
\end{align}
%where $\ket{0q} = \ket{0^{N-m}}\ket{q}$.
Finally unitary operators $\hat{U}^\dagger$ yield
\begin{equation}
    2^{-\frac{m}{2}} \sum_{q \in \{0,1\}^m} \ket{q}(\hat{U}^\dagger \ket{0q})=\ket{\phi_\mathrm{out}}.
\end{equation}
When measuring the qubits from $m+1$ to $N+m$, the probability distribution is given by
\begin{align}
    P(\sigma; \hat{U})=&\mathrm{Tr} [(\hat{I}^m \otimes \ket{\sigma} \bra{\sigma}) \ket{\phi_\mathrm{out}} \bra{\phi_\mathrm{out}}] \nonumber\\
    =& 2^{-m} \sum_q|\bra{\sigma} \hat{U}^\dagger\ket{0q}|^2,
\end{align}
which reproduces \eqref{eq:E_UO}:
\begin{align}
    \sum_\sigma \bra{\sigma} \hat{H}_\mathrm{cs} \ket{\sigma} P(\sigma; \hat{U}) =&2^{-m}\sum_q \bra{0q} \hat{U} \hat{H}_\mathrm{cs} \hat{U}^\dagger \ket{0q}\nonumber\\
    =&E(\hat{U},\hat{H}_\mathrm{cs}).
\end{align}
It indicates that $E(\hat{U},\hat{H}_\mathrm{cs})$ is efficiently computed by gate-based quantum devices using $N+m$ qubits.

    \label{expectation}
%\end{align}
%which reproduces \eqref{eq:E_UO}.
%It indicates that $E(\hat{U},\hat{H}_\mathrm{cs})$ is efficiently computed by gate-based quantum devices using $2N+1$ qubits.

%\section{Max-$4$ cut results}

\section{Details of the noise simulation}
In our simulation, we introduce independent depolarizing channels on qubits~$i$ and~$j$ each time a two-qubit gate such as $\mathrm{CX}_{i,j}$ is applied.
We assume that two-qubit gates can be applied between any pair of qubits.
Other sources of noise, such as single-qubit gate errors, readout errors, the initial-state preparation errors, are neglected.
%We decompose the gate $\exp(-i \gamma \hat{\sigma}_j^z \hat{\sigma}_k^z)$ in $U_\mathrm{C}(\gamma)$ into $\mathrm{CX}_{j,k} \exp(-i \gamma \hat{\sigma}_k^z) \mathrm{CX}_{j,k}$

The noisy two-qubit gate is represented by the superoperator on the density matrix $\rho$,
\begin{equation}
    \mathcal{D}_{i,j} [\rho]=D_i \otimes D_j [\mathrm{\Lambda}_{i,j} \rho \mathrm{\Lambda}_{i,j}],
\end{equation}
where $\Lambda_{i,j}$ is a two-qubit gates acting on qubits~$i$ and~$j$, $D_i$ denotes the depolarizing noise acting on qubit~$i$, defined as
\begin{equation}
    D_i [\rho] =(1-p)\rho + \frac{p}{3} \sum_{\alpha \in \{x,y,z\}} \hat{\sigma}_i^\alpha \rho \hat{\sigma}_i^\alpha.
\end{equation}
Here $p$ is the noise strength, assumed to be uniform across all qubits.
The two-qubit gate error rate is defined as
\begin{align}
    \epsilon=&1-[\bra{\psi} \Lambda_{i,j} (D_{i,j}[\ket{\psi}\bra{\psi}]) \Lambda_{i,j} \ket{\psi}]_\psi \nonumber\\
    =&\frac{4p}{5}(2-p),
\end{align}
where $[\cdot]_\psi$ denotes the ensemble average over a state of $2$-qubits $\ket{\psi}$ drawn from the uniform (Haar) measure~\cite{nielsen2002simple}.

To numerically implement the depolarizing noise channel $D_i[\rho]$, we employ a stochastic unraveling method.
Instead of simulating the density matrix $\rho$, we simulate the stochastic dynamics of a pure state $\ket{\psi}$ and perform an ensemble average to recover the mixed-state dynamics (i.e., $\rho=\overline{\ket{\psi} \bra{\psi}}$, where the overline indicates averaging).
Concretely, for each application of the depolarizing noise channel, we apply a unitary rotation:
\begin{equation}
    \ket{\psi'}=e^{\mathrm{i} \xi \vec{\sigma}_i \cdot \vec{n}} \ket{\psi},
\end{equation}
where $\vec{n}=(n_x, n_y, n_z)$ is a uniformly sampled unit vector.
The resulting averaged density matrix becomes
\begin{equation}
    \overline{\ket{\psi'}\bra{\psi'}}=\cos^2 \xi \overline{\ket{\psi}\bra{\psi}} + \frac{\sin^2 \xi}{3} \sum_{\alpha \in \{x,y,z\}} \hat{\sigma}_i^\alpha \overline{\ket{\psi}\bra{\psi}} \hat{\sigma}_i^\alpha,
\end{equation}
which reproduces the depolarizing channel with strength $p=\sin^2 \xi$.
In the numerical simulation, we average over $10$~independent samples to approximate the ensemble average.

\section{Compression unitary operators for typical inequality constraints}
\begin{algorithm}[t]
\caption{$\hat{U}_\mathrm{cs}^\dagger$ construction using the D-ansatz} \label{Algo:D-ansatz}
\begin{algorithmic}[1]
%% Input / Output
\renewcommand{\algorithmicrequire}{\textbf{Input:}}
\Require Number of qubits in original space $N$, Number of qubits in compressed space $m$, Number of layers $\ell$, and Binary parameters $ \{x^{(k)}\} \in \{0,1\}$
\renewcommand{\algorithmicensure}{\textbf{Output:}}
\Ensure $\hat{U}_\mathrm{cs}^\dagger$
\State $\hat{U} \gets \hat{I}^{N}$ and $k=0$
\For{$i=1$ to $N-m$}
    \State $k \gets k + 1$
    \If{$x^{(k)}=1$}
        \State $\hat{U} \gets \hat{\sigma}_i^x \hat{U}$
%                        \State $x_j \gets 1-x_j$
    \EndIf
    %\State Update $m^{(k)}$ to $m^{(k+1)} (\leq m^{(k)})$
    %\State Generate $\hat{U}'$ for the $k+1$th constraint %using a deterministic method or a variational method
    %\State $\hat{U}^{(k+1)} \gets \hat{U}' \hat{U}^{(k)}$
\EndFor
\For{$i_\ell=1$ to $\ell$ }
    \For{$i_1, i_2,i_3 \in [1,N] (i_1 \neq i_2, i_1 \neq i_3, i_2 < i_3)$}
        %\If{$i_1\neq i_2$ and $i_1 \neq i_3$}
        \State $k \gets k+1$
        \If{$x^{(k)}=1$}
            \State $\hat{U} \gets \mathrm{CSWAP}_{i_1,i_2,i_3} \hat{U}$
%                        \State $x_j \gets 1-x_j$
        \EndIf
        %\EndIf
    \EndFor
    \For{$i_1, i_2 \in [1,N] (i_1 \neq i_2)$}
        \State $k \gets k+1$
        \If{$x^{(k)}=1$}
            \State $\hat{U} \gets \mathrm{CX}_{i_1,i_2} \hat{U}$
%                        \State $x_j \gets 1-x_j$
        \EndIf
    \EndFor
    \For{$i \in [1,N]$}
        \State $k \gets k+1$
        \If{$x^{(k)}=1$}
            \State $\hat{U} \gets \hat{\sigma}_i^x \hat{U}$
%                        \State $x_j \gets 1-x_j$
        \EndIf
    \EndFor
\EndFor
\State $\hat{U}_\mathrm{cs}^\dagger \gets \hat{U}$
%¥For{$m = 1$ to $M$}
%        \State $f \gets \mathsf{True}$
%        \While{$f = \mathsf{True}$}
%                \For{$i \in V$}
                        %¥State $¥Delta Q_i ¥gets$ Energy change of $Q'$ ¥par
                        %hskip¥algorithmicindent ¥hspace{15mm} when flipping $x_i$
%                        \State $\Delta Q_i \gets$ Energy change of $Q'$ when flipping $x_i$
%                \EndFor
%                \State $j \gets \min(\argmin_{i \in V} \Delta Q_i$)
%                \If{$\Delta Q_j < 0$}
%                        \State $x_j \gets 1-x_j$
%                \Else
%                        \State $f \gets \mathsf{False}$
%                \EndIf
%        \EndWhile
%¥EndFor
%\State $\sigma_i \gets 2 x_i - 1$ for $i \in V$
\end{algorithmic}
\end{algorithm}

\begin{table*}[t]
  \centering 
  \caption{List of inequalities and compression unitary operators} 
  \scalebox{1}{
  \begin{tabular}{ll}\hline
Constraint: $0 \leq \sum_{i=1}^N x_i \leq 1$ & Constraint: $0 \leq \sum_{i=1}^N x_i \leq 1$\\
%Weight: w[1] = 1, w[2] = 1, w[3] = 1\\
Number of qubits in original space $N$: 3 & Number of qubits in original space $N$: 4\\
Number of qubits in compressed space $m$: 2 & Number of qubits in compressed space $m$: 3\\
FS ratio in original space: 0.5 & FS ratio in original space: 0.3125\\
FS ratio in compressed space: 1.0 & FS ratio in compressed space: 0.625\\
Label: [4 2 0 1] & Label: [ 8  0 12  2  6  1  4 10]\\
Number of layers $\ell$: 1 & Number of layers $\ell$: 1\\
Parameters: [0 0 1 0 1 1 0 1 0 0 0 0 1] & Parameters: [0 0 1 1 1 1 0 0 0 1 0 1 1 0 0 0 0 1 1 0 0 0 0 1 1 0 0 0 1]\\
Number of parameters: 6 & Number of parameters: 12\\
Number of CSWAP gates: 2 & Number of CSWAP gates: 7\\
Number of CNOT gates: 3 & Number of CNOT gates: 4\\
Number of X gates: 1 & Number of X gates: 1\\ \hline
Constraint: $0 \leq \sum_{i=1}^N x_i \leq 1$ & Constraint: $0 \leq \sum_{i=1}^N x_i \leq 2$\\
Number of qubits in original space $N$: 5 & Number of qubits in original space $N$: 5\\
Number of qubits in compressed space $m$: 3 & Number of qubits in compressed space $m$: 4\\
FS ratio in original space: 0.1875 & FS ratio in original space: 0.5\\
FS ratio in compressed space: 0.75 & FS ratio in compressed space: 1.0\\
Label: [ 4  8  6  1  0 10  2 16] & Label: [ 5  1  8  2  3 17 10 24  0 16  6 18 20 12  4  9]\\
Number of layers $\ell$: 1 & Number of layers $\ell$: 2\\
Parameters: [0 1 1 1 1 1 0 0 0 0 0 0 1 1 1 0 0 0 0 0 0 0 0 1 1 0 1 1 0 0 0 0 & Parameters: [1 0 0 1 0 0 1 0 0 0 1 0 0 0 0 1 0 1 1 0 1 0 0 1 0 1 0 1 0 1 1 0\\
1 0 1 0 0 1 0 1 1 0 1 0 0 1 1 1 1 0 0 0 1 0 1 1 0] & 1 0 0 0 1 1 1 1 1 0 0 1 0 0 1 1 0 0 1 1 1 0 0 0 0 0 0 0 1 1 0 0 0 0 0 0 0 0 1\\
& 0 0 1 0 1 0 0 0 1 1 1 1 0 1 1 0 0 0 1 0 0 0 1 0 0 0 1 0 0 1 0 1 0 1 1 0 0 1 0\\
& 0]\\
Number of parameters: 25 & Number of parameters: 44\\
Number of CSWAP gates: 11 & Number of CSWAP gates: 23\\
Number of CNOT gates: 10 & Number of CNOT gates: 17\\
Number of X gates: 4 & Number of X gates: 4\\ \hline
Constraint: $0 \leq \sum_{i=1}^N x_i \leq 1$ & Constraint: $0 \leq \sum_{i=1}^N x_i \leq 2$\\
Number of qubits in original space $N$: 6 & Number of qubits in original space $N$: 6\\
Number of qubits in compressed space $m$: 3 & Number of qubits in compressed space $m$: 5\\
FS ratio in original space: 0.109375 & FS ratio in original space: 0.34375\\
FS ratio in compressed space: 0.875 & FS ratio in compressed space: 0.6875\\
Label: [ 0 32  2 16  4 34  1  8] & Label: [41 36 24 32 44 50 40  5 16  9  0 22 20 12  2 19 48 33 18 10  8 34 23\\
&1 7 37  4 21  3 14  6 17]\\
Number of layers $\ell$: 1 & Number of layers $\ell$: 1\\
Parameters: [1 0 0 0 1 0 0 1 0 0 0 1 1 0 0 0 0 0 0 0 0 1 0 1 0 0 1 1 0 0 1 0 & Parameters: [1 0 0 0 0 0 1 0 1 1 0 1 0 1 1 1 1 1 0 0 0 0 1 0 0 0 1 0 1 0 1 0\\
1 0 0 1 0 0 0 1 0 1 0 0 1 0 1 1 1 0 0 0 0 1 0 0 1 0 1 1 1 0 0 1 1 0 1 0 1 0 0 & 1 0 0 0 0 0 0 0 0 0 0 1 0 0 1 0 1 1 1 1 0 1 0 0 1 1 1 1 0 1 1 0 1 0 1 1 0 1 0\\
1 1 1 0 0 0 0 1 1 0 1 0 1 1 0 0 0 0 0 0 0 1 1 1 1 0 1 1] & 0 0 0 0 0 0 0 0 1 0 1 1 1 0 0 0 0 1 0 0 1 1 0 1 0 1]\\
Number of parameters: 41 & Number of parameters: 41\\
Number of CSWAP gates: 22 & Number of CSWAP gates: 25\\
Number of CNOT gates: 13 & Number of CNOT gates: 11\\
Number of X gates: 6 & Number of X gates: 5\\ \hline
Constraint: $0 \leq \sum_{i=1}^N x_i \leq 1$ & Constraint: $0 \leq \sum_{i=1}^N x_i \leq 1$\\
Number of qubits in original space $N$: 7 & Number of qubits in original space $N$: 8\\
Number of qubits in compressed space $m$: 3 & Number of qubits in compressed space $m$: 4\\
FS ratio in original space: 0.0625 & FS ratio in original space: 0.03515625\\
FS ratio in compressed space: 1.0 & FS ratio in compressed space: 0.5625\\
Label: [16  1  2  0 32  8 64  4] & Label: [128  16   0  20  32  18   2  64  96   4   3   8  34   6   1  10]\\
Number of layers $\ell$: 1 & Number of layers $\ell$: 1\\
Parameters: [1 1 1 1 1 0 0 1 1 0 0 0 1 0 1 0 0 0 1 1 1 1 0 1 0 1 1 0 0 1 1 0 & Parameters: [0 0 0 0 1 1 1 0 1 1 1 0 0 1 0 1 1 1 0 1 0 1 0 0 1 1 1 1 1 1 0 0 \\
0 1 1 1 0 0 1 0 1 1 1 1 0 0 1 0 0 0 1 0 1 1 1 0 0 0 0 0 1 0 1 0 1 1 0 1 0 0 1 & 0 1 0 0 0
 1 1 1 0 0 1 1 0 0 0 0 1 1 0 1 0 0 0 1 0 0 1 0 0 1 0 1 1 0 0 1 0 0 0 \\
 1 1 1 1 1 0 0 1 0 0 0 0 0 1 0 1 1 1 1 0 1 0 0 1 1 0 1 0 0 1 1 1 0 1 1 1 1 0 1 & 1 0 0 1 0 0 0 1 0 0 0 1 1 0 1 1 0 0 1 1 1 1 0 1 0 0 0 0 1 0 1 1 1 1 0 1 1 0 0 \\
 1 0 0 0 0 1 1 0 0 0 0 0 0 0 0 0 0 0 0 1 0 0 0 0 0 1 0 1 1 0 0 0 0 1 0 0 0 1 0 & 0 1 1 1 0 0 1 1 0 1 0 1 1 0 0 0 1 0 1 1 0 1 1 0 0 1 0 1 0 0 0 0 1 1 1 1 0 1 1 \\
 0 0 1 0 1 1 1 0 0] & 1 1 0 1 0 1 0 0 0 1 0 0 1 1 0 0 0 0 0 1 0 1 1 1 0 0 0 0 0 0 1 0 0 0 0 0
 0 0 0\\
 & 0 1 1 0 0 1 1 0 0 1 1 0 0 0 0 0 1 0 0 0 0 0 0 0 1 0 0 0 0 0 0 0 1 0
 0 0 0 0 1\\
 & 1 0 0 0 0 0 0 0 1]\\
Number of parameters: 73 & Number of parameters: 97\\
Number of CSWAP gates: 55 & Number of CSWAP gates: 83\\
Number of CNOT gates: 10 & Number of CNOT gates: 13\\
Number of X gates: 8 & Number of X gates: 1\\ \hline
    \end{tabular}
  }
  \label{Table:sample}
\end{table*}

\begin{figure}
\begin{minipage}[]{0.95\linewidth}
\centering
%\subcaption{}
\scalebox{1}{
\begin{quantikz}
    \lstick[wires = 1]{$0$}  &\swap{1}&\ctrl{1}&\ctrl{2}&{}&{}&\rstick[wires = 1]{$x_1$}\\
    \lstick[wires = 1]{$x'_1$}  & \ctrl{1} &\targ{}&{}&\ctrl{1}&{}&\rstick[wires = 1]{$x_2$}\\
    \lstick[wires = 1]{$x'_2$}  &\targX{}&{}&\targ{}&\targ{}&\targ{}&\rstick[wires = 1]{$x_3$}
\end{quantikz}
}
\end{minipage}
\caption{$\hat{U}_\mathrm{cs}^\dagger$ for $0\leq x_1+x_2+x_3 \leq 1$.}
\label{fig:sample}
\end{figure}

We provide unitary compression operators corresponding to typical inequality constraints.
Algorithm~\ref{Algo:D-ansatz} outlines the construction of $\hat{U}_\mathrm{cs}^\dagger$ using the D-ansatz shown in Fig.~\ref{fig:compress_ansatz}~(b).
The D-ansatz takes as input the parameters $N$, $m$, $\ell$, and a binary vector $\{x^{(k)}\}$, where the $k$-th gate is activated when $x^{(k)}=1$.
The circuit consists of X gates (lines~$2$--$7$ and lines~$21$--$26$), CSWAP gates (lines~$9$--$14$), and CNOT gates (lines~$15$--$20$).
Here, $\mathrm{CSWAP}_{i_1,i_2,i_3}$ acts on qubits $i_1$, $i_2$, and $i_3$ with $i_1$ serving as the control qubit.

Table~\ref{Table:sample} presents the information to specify the corresponding unitary $\hat{U}_\mathrm{cs}^\dagger$ for the inequality constraints $\ell \leq \sum_{i=1}^N x_i \leq u$.
%In this file, the values following $\mathrm{N}$, $\mathrm{l}$, $\mathrm{u}$, and $\mathrm{w}$ represent the number of variables involved in the constraint, the lower and the upper bounds, and the weight vector, respectively.
%The constraint takes the form of .
The columns ``FS ratio'' indicate the fraction of feasible solutions in the corresponding spaces.
The ``Label'' column provides the mapping between the state in the compressed space $a'=\sum_{i=1}^{m}2^{i-1} x'_{i}$ and the state in the original space $a=\sum_{i=1}^{N}2^{i-1} x_{i}$ (i.e., $\mathrm{Label}[a']=a$).
%In the example of constraint $0\leq x_1+x_2+x_3 \leq 1$, the compressed state $(x'_1,x'_2)=(1,0)$ is mapped to the state $(x_1,x_2,x_3)=(0,1,0)$ via the unitary transformation.
The ``Parameters'' column lists the $x^{(k)}$ values used in Algorithm~\ref{Algo:D-ansatz} to define the specific quantum circuit for $\hat{U}_\mathrm{cs}^\dagger$.
For example, Fig.~\ref{fig:sample} illustrates the quantum circuit corresponding to the inequality constraint $0\leq x_1+x_2+x_3 \leq 1$. 
The table also reports the total number of parameters, CSWAP-gates, CNOT-gates, and X-gates used in the construction.

\end{appendices}

\section*{Acknowledgment}
This work was supported by JSPS KAKENHI (Grant Number 23K13034) and the New Energy and Industrial Technology Development Organization (NEDO), Japan, under Project JPNP16007.
%This work was supported in part by JST CREST Grant Number JPMJCR19K4, Japan.
%We acknowledge the use of IBM Quantum services for this work.
%The views expressed are those of the authors, and do not reflect the official policy or position of IBM or the IBM Quantum team.

\bibliographystyle{IEEEtran}

\bibliography{vqa.bib}

\EOD

\end{document}